\documentclass[11pt]{article}
\newtheorem{lemma}{Lemma}
\newtheorem{thm}{Theorem}

\def\be{\begin{equation}}                                            % ~~
\def\ee{\end{equation}}                                              %
\def\om{\omega}                                                     % ~~

\title{The structure of Lie algebras and the classification
problem for partial differential equations.}

\author{P.~Basarab-Horwath \\ \small Link\"oping University, S-581
83 Link\"oping, Sweden \thanks{e-mail: pehor@mai.liu.se} \and V.
Lahno\\ \small Pedagogical University, 2 Ostrohradskyj Street,
314000 Poltava, Ukraine\thanks{e-mail: laggo@poltava.bank.gov.ua}
\and R.~Zhdanov \\ \small Institute of Mathematics,  3
Tereshchenkivska Street, 252004 Kyiv, Ukraine\thanks{e-mail:
renat@imath.kiev.ua. Supported in part by the
Swedish Natural Sciences Research Council under grant number R-RA
521-2373/1999.}}

\date{}

\begin{document}

\maketitle

\begin{abstract}
The present paper solves completely the problem of the group
classification of nonlinear heat-conductivity equations of the
form\ $u_{t}=F(t,x,u,u_{x})u_{xx} + G(t,x,u,u_{x})$.\ We have
proved, in particular, that the above class contains no nonlinear
equations whose invariance algebra has dimension more than
five. Furthermore, we have proved that there are two, thirty-four,
thirty-five, and six inequivalent equations admitting one-, two-,
three-, four- and five-dimensional Lie algebras, respectively.
Since the procedure which we use, relies heavily upon the theory of
abstract Lie algebras of low dimension,  we give a detailed account
of the necessary facts. This material is dispersed in the literature
and is not fully available in English. After this algebraic part we give
a detailed description of the method and then we
derive the forms of inequivalent invariant evolution equations, and
compute the corresponding maximal symmetry algebras. The list of
invariant equations obtained in this way contains (up to a local change of
variables)
all the previously-known invariant evolution equations belonging
to the class of partial differential equations under study.

\end{abstract}

\renewcommand{\thesection}{\Roman{section}}
\setcounter{section}{0}

\newpage
\section*{Introduction}
\setcounter{equation}{0}

Modeling  phenomena in nature with partial differential equations
is one of the central problems of mathematical physics and applied
mathematics. One can even say that mathematical physics in its
classical form was created in order to provide a rigorous
mathematical foundation for describing different  phenomena in
physics, chemistry and biology by partial differential
equations. However, when one has to decide which differential
equation fits in the best way as a model for the process under
study, one has to select from a broad class of possible partial
differential equations. Even if one has taken into account all the
peculiarities of the process under study (which is hardly
possible!), there is still great freedom  in choosing possible
models. One of the principal criteria for choosing the partial
differential equations modeling real  processes is the {\it
symmetry selection principle}. By this we mean that from the whole
set of admissible models, those models which have the highest
symmetry should be selected. This point of view is supported by
the fact that the most successful mathematical models in
theoretical  and applied science have a rich symmetry structure.
Indeed, the basic equations of modern physics, the wave,
Schr\"odinger, Dirac and Maxwell equations are distinguished from
the whole set of partial differential equations by their Lie and
non-Lie (hidden) symmetry (see \cite{fun94} for more details on
symmetry properties of these equations).

The effectiveness of the symmetry (group-theoretical) approach to
the classification of admissible partial differential equations
relies heavily upon the availability of a constructive way of
describing transformation groups leaving invariant the form of a given
partial differential equation. This is done via the
well-known infinitesimal method developed by Sophus Lie
\cite{lie24,lie27} (see, e.g., \cite{o:}--\cite{fs1:}).  Given a
partial differential equation, the problem of investigating its
maximal  (in some sense) Lie invariance group reduces to solving
an over-determined system of linear  partial differential
equations, called the determining equations. However, if the
equation under study contains arbitrary elements (functions), then
one has to solve an intermediate classification problem. Namely,
it is necessary to describe all the possible forms of the
functions involved such that this equation admits a non-trivial
invariance group.

In principle, the classification problem is solved with the help
of  the Lie algebra approach. However, since the determining
equations involve some arbitrary functions, there is an evident
need for a modification of the basic Lie technique in order to
obtain an efficient and systematic way  of classifying these
arbitrary elements. The idea of this modification was suggested by
Sophus Lie himself. Indeed, his way of obtaining all ordinary
differential equations in one variable admitting non-trivial
symmetry algebras \cite{lie24,lie27} tells us what is to be done
in the case at hand. We should first construct all the
possible inequivalent realizations of symmetry algebras within
some class of Lie vector fields. If we succeed in doing this, then
the symmetry algebras will be specified, so that we can apply
directly Lie's infinitesimal algorithm, thus getting inequivalent
classes of invariant equations. In this way, Sophus Lie obtained
his famous classification of realizations of all inequivalent
complex Lie algebras in the plane \cite{lie24,lie27}. Recently,
Lie's classification was exploited by Olver and Heredero
\cite{ohe} in order to classify nonlinear wave equations in two
independent variables that are invariant with respect to
transformation groups not changing  the temporal variable.

A systematic implementation of these ideas for partial
differential equations has been worked out by Ovsjannikov
\cite{o:}. His approach is based on the concept of an equivalence
group, which is a Lie transformation group acting in the 
extended space of independent variables, functions and their
derivatives, and preserving the class of partial differential
equations under study. It is possible to modify Lie's algorithm in
order to make it applicable for the computation of this group
\cite{o:}. Next, one constructs the optimal system of subgroups of
the equivalence group. The last step uses Lie's algorithm for
obtaining specific partial differential equations, that\ (a)\
belong to the class under study, and\ (b)\ are invariant with
respect to  the subgroups mentioned above. This approach has been
applied to a number of equations of nonlinear gas dynamics and
diffusion equations (Akhatov, Gazizov and Ibragimov
\cite{ah:,ah1:}). Ovsjannikov's ideas have also been exploited by
Torrisi and co-workers in order to perform a preliminary
group classification of some nonlinear diffusion and heat
conductivity equations \cite{tor,tor1}. Ibragimov and  Torrisi
have obtained a number of important results on the group
classification of nonlinear  detonation equations \cite{ibr} and
nonlinear hyperbolic type equations \cite{ibr1}. There is a number
of papers (see, e.g., \cite{king} and the references therein)
devoted to a direct computation of equivalence groups of some
PDEs. Since the transformations of the equivalence group are used
in their finite form, this approach has the merit of giving the
possibility of finding {\em discrete} equivalence groups or even
{\it non-local} ones.

However, the possibility of implementing  Ovsjannikov's approach
in its full generality presupposes that we are able to construct
the optimal system of subgroups of the equivalence group. So that,
even for the case when the equivalence group has a finite number of parameters,
there arise major algebraic  difficulties, since for a number of
known finite-parameter Lie groups the classification problem has
not yet been solved (to say nothing about infinite-parameter  Lie
groups, where this problem is completely open). Consequently,
there is an  evident need for Ovsjannikov's approach to be
modified so that it can be applied to  the case of infinite-parameter
equivalence groups.

In the paper \cite{zhd99} we have developed a new approach that
enables us to solve efficiently the symmetry classification
problem for partial differential equations even for the case of
infinite-dimensional equivalence groups. It is mainly based on the
following  facts:
\begin{itemize}

\item If the partial differential equation possesses non-trivial
symmetry, then it is invariant under some finite-dimensional  Lie
algebra of differential operators which is completely determined
by its structural constants. In the event that the maximal algebra
of invariance is infinite-dimensional, then it  contains, as a
rule, some finite-dimensional Lie algebra.

\item If there are local non-singular changes of variables which
transform a given differential equation into another, then the
finite-dimensional Lie algebra of invariance of these equations
are isomorphic, and in the group-theoretic analysis of
differential equations such equations are considered to be
equivalent.

\end{itemize}

What we have suggested in \cite{zhd99} is a preliminary
classification of inequivalent realizations of low-dimensional Lie
algebras within some specific class of first-order linear
differential operators. This class is determined by the structure
of the equation under study. Its elements form a representation
space for realizations of Lie algebras of symmetry groups admitted
by  the equations belonging to the class of partial differential
equations under study. A natural equivalence relation is
introduced on the set of all possible realizations. Namely, two
realizations are called  equivalent if they are transformed into
each other by the action of the equivalence group.  In other
words, solving the problem of symmetry classification  of partial
differential equations  having some prescribed form, is equivalent
to constructing a representation theory of Lie transformation
groups (or Lie algebras of first-order differential operators)
realized as symmetry groups (algebras) of the equations in
question.

The first aim of the present paper is to give a detailed
exposition of our approach. A full understanding of the techniques
applied requires some basic facts from the general theory of Lie
groups and algebras, some of which are dispersed in the literature and are
not available in English (this is the case for
the papers of Mubarakzyanov and Morozov). So, in addition to the
exposition of the classification results, we give a survey of
results on the structure of Lie algebras (with special emphasis on
low-dimensional Lie algebras), which are of vital importance for
the effective implementation of our approach.

The second aim is obtaining a complete description of the nonlinear
heat conductivity equations of the form
\begin{equation}
u_{t}=F(t,x,u,u_{x})u_{xx} + G(t,x,u,u_{x}) \label{2.1}
\end{equation}
that admit non-trivial symmetry group. Hereafter $u=u(t,x)$,\
$F$,\ $G$ are sufficiently smooth functions of the corresponding
arguments,\ $u_{t}=\displaystyle{\frac {\partial u}{\partial
t}}$,\ $u_{x}=\displaystyle\frac{\partial u} {\partial x}$,
$u_{xx}=\displaystyle{\partial^{2}u \over \partial  x^{2}}$, $F
\not =0$.

Note that the above equation is, in some sense, the most general
evolution equation in one dimension. Indeed, any equation of the
most general form
\begin{equation}
u_{t}=H(t,x,u,u_{x},u_{xx}) \label{2.1g}
\end{equation}
which admits at least one-parameter symmetry group, not changing
the temporal variable, can be reduced to the form (\ref{2.1}) by a
non-point transformation. So that our group classification of
equations (\ref{2.1}) will also cover invariant equations of the
form (\ref{2.1g}) excepting for the small subclass of equations
whose symmetry algebras are spanned by operators with
non-vanishing coefficients by ${\displaystyle \partial\over
\displaystyle
\partial t}$.

The principal scheme of the paper is as follows. Section I
contains a general description of our approach. In the next
section we give a brief overview of the necessary facts from the
general theory of Lie algebras. Section III is devoted to group
classification of PDEs (\ref{2.1}). We consider subsequently, the
cases of semi-simple, semi-direct sum of semi-simple and solvable
and solvable symmetry algebras thus getting the full solution of
the classification problem for nonlinear heat conductivity
equations belonging to the class (\ref{2.1}). The last section
contains discussion of the results obtained and some conclusions.

\section{Description of the method}
\setcounter{equation}{0}

The approach to the classification of partial differential
equations which we propound is, in fact, a synthesis of Lie's
infinitesimal method, the use of equivalence transformations and
the theory of classification of abstract finite-dimensional Lie
algebras. It constitutes a constructive solution of the problem
the group classification of partial differential equations
possessing large classes of arbitrary elements and admitting {\it
non-trivial finite-dimensional} invariance algebras.

The realization of group classification in the proposed approach
consists in the implementation of the following algorithm:

\begin{itemize}

\item[I] The first step involves finding the form of the infinitesimal
operators which generate the symmetry group of the  equation under
consideration, and the construction of the equivalence group of
this equation. To find the form of the infinitesimal operators one
uses the usual Lie algorithm. As a result we obtain a system of
linear partial differential equations of first order, which
connect the coefficients of the infinitesimal operators with the
arbitrary term of the equation. In what follows, we call this
system the characterizing system of the equation. In order to
construct the equivalence group $\cal E$ of the equation under
consideration, one can use the infinitesimal as well as the direct
method.

\item[II] In the second step, one carries out the group
classification of those equations of the given form which admit
finite-dimensional Lie algebras of invariance.

For this, one carries out a step-by-step classification of
finite-dimensional Lie algebras within the specified class of
infinitesimal operators, up to equivalence under transformations
of the group $\cal E$. In this, one has to see if each algebra
obtained in this way can be an invariance algebra of the equation
at hand before proceeding from the realization of Lie algebras of
lower dimension to the realization of Lie algebras of higher
dimension. This eliminates superfluous realizations of Lie
algebras. Also, those realizations of Lie algebras which are
invariance algebras of the equation will, as their dimension
increases, correspond to greater fixing of the arbitrary term.

This procedure is continued until the arbitrary term in the
equation is completely determined or until it is no longer
possible  to extend the realization of Lie algebras beyond a given
dimension within the specified class of infinitesimal operators.

\item[III] The third step is then to exploit the characterizing system
or the infinitesimal method of Lie in order to find, for  each of
the particular choices of the arbitrary term, the maximal
invariance algebra of the equation under consideration.
Furthermore, the equivalence of the equations obtained in this
manner is determined. We note that, in as much as equivalent
equations have isomorphic invariance algebras, we may test the
realizations of the invariance algebras for equivalence rather
than the equations themselves.

\end{itemize}

Note that similar ideas have been used by Gangon and Winternitz
\cite{win} in order to classify symmetries of nonlinear
Schro\"odinger equations having variable coefficients.

\section{Lie-algebraic structures involved in the classification
algorithm} \setcounter{equation}{0} \setcounter{thm}{0}
\setcounter{rmk}{0} \setcounter{lemma}{0}

Let us take a more detailed look at the second step of the
algorithm. As is clear from what has been said above, carrying out
this step assumes that there is a classification of non-isomorphic
finite-dimensional Lie algebras (in particular, we are interested
in a classification of Lie algebras over the real numbers).

One of the central theorems which deals with the structure of Lie
algebras is the Levi-Mal'cev theorem:

\begin{thm} Let $L$ be a finite-dimensional Lie algebra over ${\bf R}$
or ${\bf C},$ and let $N$ denote its radical (the largest solvable
ideal in $L$). Then there exists a semi-simple Lie subalgebra $S$
of $L$ such that

\begin{equation}\label{eq1}
L=S\subset \hskip -3.8mm + N
\end{equation}
\end{thm}
Equation (\ref{eq1}) is called the Levi decomposition of the Lie
algebra $L,$ and the semi-simple subalgebra $S$ is called the Levi
factor.

The Levi-Mal'cev decomposition gives us

\[
[N,N]\subset N,\;\; [S,S]\subset S,\;\; [N,S]\subset N,
\]
so that any Lie algebra $L$ is the semi-direct sum $\displaystyle
L=S\subset \hskip -3.8mm + N$ of its maximal solvable ideal $N$
and the semi-simple subalgebra $S.$ We see then that this result
reduces the task of classifying all Lie algebras to the following
problems:

\begin{itemize}

\item[1)] the classification of all semi-simple Lie algebras;
\item[2)] the classification of all solvable Lie algebras;
\item[3)] the classification of all algebras which are semi-direct
sums of semi-simple Lie algebras and solvable Lie algebras.
\end{itemize}

\subsection{Semi-simple Lie algebras.}

Of the problems listed above, only that of classifying all
semi-simple Lie algebras is completely solved. We have the
well-known theorem due to Cartan:

\begin{thm}({\bf Cartan's theorem})
Any semi-simple complex or real semi-simple Lie algebra can be
decomposed into a direct (Lie algebra) sum of ideals which are
mutually orthogonal simple subalgebras. Here, orthogonality is
with respect to the Cartan-Killing form $\displaystyle
(X,Y)=Tr(ad\,X,ad\,Y).$ \end{thm}

Let $L$ be a semi-simple Lie algebra. Then, by Cartan's theorem,
we have

\[
L=S_1\oplus S_2\oplus\ldots\oplus S_m,
\]
where $S_1,\ldots,S_m$ are simple Lie algebras. Thus, the problem
of classifying semi-simple Lie algebras is equivalent to that of
classifying all non-isomorphic simple Lie algebras. This
classification is known (see, for instance, \cite{br}).

There are four sequences of classical Lie algebras $A_n (n\geq
1),\, B_n (n\geq 1),\, C_n (n\geq 1),\, D_n (n\geq 1)$ and five
exceptional Lie algebras $G_2,\, F_4,\, E_6,\, E_7,\,E_8$ which
together exhaust all the simple complex Lie algebras. There are
some isomorphisms between some of these algebras. Indeed
\cite{hel} there are the following isomorphisms:

\[
A_1\cong B_1\cong C_1,\;\; B_2\cong C_2,\;\; A_3\cong D_3,\;\;
D_2\cong A_1\oplus A_1
\]
and there are no other isomorphisms between the series.

The dimensions of the classical complex Lie algebras $A_n,\,
B_n,\, C_n$ and $D_n$ are given in the following table:
\vspace{2mm}

\begin{tabular}{|c|c|c|c|c|}\hline
 Algebra & $A_n$ & $B_n$ & $C_n$ & $D_n$\\ \hline
Dimension & $n(n+2)$ & $n(2n+1)$ & $n(2n+1)$ & $n(2n-1)$\\ \hline
\end{tabular}

\vspace{2mm}
 The dimensions of the exceptional Lie algebras are
all even: $\dim G_2=14,\; \dim F_4=52,\; \dim E_6=78,\; \dim
E_7=133,\; \dim E_8=248.$

To describe the real simple Lie algebras one uses the fact that
every simple Lie algebra over the reals ${\bf R}$ is either a
simple algebra over the complex field ${\bf C}$ (considered as an
algebra over ${\bf R}$), or it is the real form of a simple Lie
algebra over ${\bf C}.$

The real classical Lie algebras play an important role in the
group analysis of differential equations. Below, we give a more
detailed description of these Lie algebras. The symbol $L_k$
denotes a compact simple Lie algebra.

\medskip
\begin{itemize}

\item[I.]\underline{Real forms of the algebras $sl(n,{\bf C})\,(\cong
A_{n-1},\, n\geq 2)$}

\item[1)] $L_k=su(n),$ the Lie algebra of all skew-symmetric matrices
$Z$ of order $n$ with  $Tr\, Z=0$ of order $n$ with $Tr\, Z=0.$

\item[2)] $sl(n,{\bf R}),$ the Lie algebra of all real matrices $X$  of
order $n$ with  $Tr\, X=0.$

\item[3)] $su(p,q),\, p+q=n,\, p\geq q,$ the Lie algebra of all matrices
of the form

\[
\left[\begin{array}{cc} Z_1 & Z_2\\Z^{*}_2 & Z_3\end{array}
\right]
\]
where $Z_1,\, Z_3$ are skew-symmetric matrices of order $p$ and
$q$ respectively, $Tr(Z_1+Z_3)=0,$ and $Z_2$ is an arbitrary
matrix of order $q.$

\item[4)] $su^{*}(2n),$ the Lie algebra of all complex matrices of order
$2n$ of the form
\[
\left[\begin{array}{cc} Z_1 & Z_2\\-\overline{Z}_2
&\overline{Z}_1\end{array} \right]
\]
where $Z_1,\, Z_2$ are complex matrices of order $n$ with
$Tr(Z_1+\overline{Z}_1)=0.$

\item[II.]\underline{Real forms of the algebras $so(2n,{\bf C})\,(\cong
D_{n},\, n\geq 1)$}

\item[1)] $L_k=so(2n),$ the Lie algebra of all real skew-symmetric
matrices of order $2n.$

\item[2)] $so(p,q),\, p+q=2n,\, p\geq q,$ the Lie algebra of all real
matrices of order $2n$ of the form
\[
\left[\begin{array}{cc} X_1 & X_2\\X^{T}_2 & X_3\end{array}
\right]
\]
where all the $X_i$ are real matrices, and  $X_1,\, X_3$ are
skew-symmetric matrices of order $p$ and $q$ respectively, and
$X_2$ is an arbitrary matrix of order $q.$

\item[3)] $so^{*}(2n),$ the Lie algebra of all complex matrices of order
$2n$ of the form

\[
\left[\begin{array}{cc} Z_1 & Z_2\\-\overline{Z}_2
&\overline{Z}_1\end{array} \right]
\]$Z_1$ skew-symmetric and $Z_2$ Hermitian.

\item[III.]\underline{Real forms of the algebras $so(2n+1,{\bf
C})w\,(\cong B_{n},\, n\geq 1)$}

\item[1)] $L_k=so(2n+1),$ the Lie algebra of all real skew-symmetric
matrices of order $2n+1.$

\item[2)] $so(p,q),\, p+q=2n+1,\, p\geq q,$ the Lie algebra of all real
matrices of order $2n+1$ of the form

\[
\left[\begin{array}{cc} X_1 & X_2\\X^{T}_2 & X_3\end{array}
\right]
\]
where all the $X_i$ are real matrices, and  $X_1,\, X_3$ are
skew-symmetric matrices of order $p$ and $q$ respectively, and
$X_2$ is an arbitrary matrix of order $q.$

\item[IV.]\underline{Real forms of the algebras $sp(n,{\bf C})\,(\cong
C_{n},\, n\geq 1)$}

\item[1)] $L_k=sp(n),$ the Lie algebra of all matrices of order $2n$ of
the form

\[
\left[\begin{array}{cc} Z_1 & Z_2\\Z_3 &-Z^{T}_1\end{array}
\right]
\]
where all the $Z_i$ are complex matrices of order $n$ and $Z_2,\,
Z_3$ are symmetric.

\item[2)] $sp(n,{\bf R}),$ the Lie algebra of all real matrices of order
$2n$ of the form

\[
\left[\begin{array}{cc} X_1 & X_2\\X_3 & -X^{T}_1\end{array}
\right]
\]
where $X_1,\, X_2,\, X_3$ are all real matrices of order $n,$ and
$X_2,\, X_3$ are symmetric.

\item[{3)}] $sp(p,q),\, p+q=n, p\ge q,$ the Lie algebra of all
complex matrices of order $2n$ of the form

\[
\left[\matrix {Z_{11}& Z_{12}& Z_{13}& Z_{14} \cr Z^{*}_{12}&
Z_{22}& Z^T_{14}& Z_{24}\cr -\overline{Z}_{13}& \overline{Z}_{14}&
\overline{Z}_{11}&-\overline{Z}_{12}\cr Z^{*}_{14}&
-\overline{Z}_{24}& -\overline{Z}^T_{12}& \overline{Z}_{22}\cr
}\right]
\]
where the  $Z_{ij}$ are complex matrices, $Z_{11}$ and $Z_{13}$
are of order $p$, $Z_{12}$ and $Z_{14}$ are $p\times q$ matrices,
$Z_{11}$ and $Z_{22}$ are skew-Hermitian, and $Z_{13}$ and
$Z_{24}$ are symmetric.
\end{itemize}

\medskip
The structure of the above real, simple classical Lie algebras is
such that every algebra of a higher dimension contains, as a
subalgebra, an algebra of the same class but of lower dimension.
This allows us to proceed  step-by-step when we study the
realizations of these algebras as vector fields, at each stage
extending the realizations of lower dimension to realizations of
higher dimension. If at some stage in this procedure the chain
stops, then this implies that there are no realizations within the
given type of vector fields of Lie algebras of higher dimension.

In searching for realizations of the classical simple Lie algebras
over ${\bf R},$ it is important to take into account the
isomorphisms for the lower-dimensional classical Lie algebras:

\begin{eqnarray*}
&& su(2) \cong so(3) \cong sp(1); \\ && sl(2,{\bf R}) \cong
su(1,1) \cong so(2,1) \cong sp(1,{\bf R});\\ && so(5) \cong
sp(2);\\ && so(3,2) \cong sp(2,{\bf R}); \\ && so(4,1) \cong
sp(1,1); \\ && so(4) \cong so(3) \oplus so(3)\cong sp(1)\oplus
sp(1); \\ && so(5) \cong sp(2);\\ && so(2,2) \cong sl(2,{\bf R})
\oplus sl(2,{\bf R});\\ && sl(2,C) \cong so(3,1); \\ && su(4)
\cong so(6); \\ && sl(4,{\bf R}) \cong so(3,3); \\ && su(2,2)
\cong so(4,2); \\ && su(3,1) \cong so^{*}(6); \\ && su^{*}(4)
\cong so(5,1);\\ && so^{*}(8)\cong so(6,2);\\ && so^{*}(4) \cong
su(2) \oplus sl(2,{\bf R}).
\end{eqnarray*}

It is not difficult to see that the Lie algebras of the first two
rows have the lowest dimension $n=3.$ Thus, in constructing
realizations of the classical simple Lie algebras over ${\bf R}$
one may begin with the algebras
\[
so(3) = \langle e_1, e_2, e_3 \rangle, \ [e_1, e_2] = e_3, [e_2,
e_3] = e_1, [e_3, e_1]=e_2;$$ $$sl(2,{\bf R}) = \langle e_1, e_2,
e_3 \rangle, [e_1, e_3] = -2 e_2, \ [e_1, e_2] = e_1, \ [e_2, e_3]
= e_3.
\]

\medskip
Maximal compact subalgebras play an important role for the
structure of the simple (and semi-simple) Lie algebras. We have:

\begin{thm}({\bf Cartan's Theorem}) A semi-simple real Lie algebra has
a decomposition of the form

\begin{equation}\label{2}
L = K \dot{+} P,
\end{equation}
where

\begin{equation} \label{3}
[K,K] \subset K, \ \ \ [K,P] \subset P, \ \ \ [P,P] \subset K,
\end{equation}
and

\begin{eqnarray} \label{4}
&& (X,X)<0 {\rm \ \ for \ \ } X \not =0 {\rm \ \ in\ \ } K,
\nonumber \\ && (Y,Y)>0 {\rm \ \ for \ \ } Y \not =0 {\ \ \rm in\
\ } P.
\end{eqnarray}

If the conditions (\ref{3}), (\ref{4}) are satisfied, then $K$ is
a maximal compact subalgebra of $L.$
\end{thm}

The decomposition (\ref{2}) for a real semi-simple Lie algebra is
called the {\it Cartan decomposition.}

Consider as an example $so(3,1),$which is the Lie algebra of the
Lorentz group. Denoting by $K_i \ (i = 1, 2, 3)$ the generators of
the compact algebra $so(3)$, and by $N_i \ (i=1,2,3)$ the
generators of the Lorentz boosts, we obtain the commutation
relations

\begin{eqnarray}
[K_i, K_j]&=& \varepsilon_{ijl} K_l, \nonumber \\ \lbrack K_i, N_j
\rbrack &=& \varepsilon_{ijl} N_l, \nonumber \\ \lbrack  N_i, N_j
\rbrack &=& -\varepsilon_{ijl} K_l.\nonumber
\end{eqnarray}
Thus, when looking for realizations of the Lie algebra $so(3,1),$
one may use the realizations obtained for the Lie algebra $so(3).$

In the table below we give the maximal compact subalgebras of the
real classical Lie algebras which are non-compact:
\vspace{2mm}

\begin{center}
\begin{tabular}{|c|c|c|c|c|c|} \hline
No/o& $L$&$K$&No/o& $L$&$K$ \\ \hline 1& $sl(n,{\bf R})$&$so(n)$&
5& $so^{*}(2n)$&$u(n)$ \\ \hline 2&$su(p,q)$&$s(u(p)\oplus
u(q))$&6& $sp(n,{\bf R})$&$u(n)$ \\ \hline 3&$
su^{*}(2n)$&$sp(n)$&7&$sp(p,q)$&$sp(p) \oplus sp(q)$ \\ \hline 4&$
so(p,q)$&$so(p) \oplus so(q)$& & & \\ \hline
\end{tabular}
\end{center}
\vspace{2mm}
Here, $u(n)$ is the Lie algebra of the unitary group $U(n),$ and
$s(u(p)\oplus u(q))$ is the set of all elements $x\in u(p)\oplus
u(q)$ such that $Tr\, x=0.$ Note that the matrix $e_{ij},$ defined
as a matrix of order $n$ with $1$ in the $(i,j)$ position and
zeroes in all other entries, is an element of $u(n).$

Because of their large dimension, the exceptional Lie algebras do
not play as important a role as the classical simple Lie algebras
do in the group analysis of differential equations. For this
reason, we only mention briefly the real forms of the algebras of
the type $G_2,\, F_4,\, E_6,\, E_7,\, E_8,$ and we consider those
subalgebras whose realizations one may use for the construction of
realizations of the real exceptional simple Lie algebras. More
details about these algebras can be found in \cite{hel}.

The algebra $G_2$ has real compact form $g_2$ and one real
non-compact form $g^{'}_2$. Moreover, $g_2 \cap g^{'}_2 \cong
su(2) \oplus su(2)$.

The algebra $F_4$ has real compact form $f_4$ and two real
non-compact forms $f^{'}_4, f^{''}_4$.  We also have $f^{'}_4\cap
f_4 \cong sp(3) \oplus su(2), f^{''}_4 \cap f_4 \cong so(9).$

The algebra $E_6$ has real compact form $e_6$ and four real
non-compact forms $e^{'}_6$, $e^{''}_6$, $e^{'''}_6$, $e^{IV}_6$.
Moreover, $e^{'}_6 \cap e_6 \cong sp(4),$ $e^{''}_6 \cap e_6 \cong
su(6) \oplus su(2)$, $e^{'''}_6 \cap e_6 \cong so(10) \oplus {\bf
R}$, $e^{IV}_6 \cap e_6 \cong f_4$.

The algebra $E_7$ has real compact form $e_7$ and four real
non-compact forms $e^{'}_7$, $e^{''}_7$, $e^{'''}_7$. We also have
$e^{'}_7 \cap e_7 \cong su(8)$, $e^{''}_7 \cap e_7 \cong so(12)
\oplus su(2)$, $e^{'''}_7 \cap e_7 \cong e_6 \oplus {\bf R}$.

The algebra $E_8$  has real compact form $e_8$ and two real
non-compact forms $e^{'}_8, e^{''}_8$. Also, $e^{'}_8 \cap e_8
\cong e_7 \oplus su(2)$, $e^{''}_8 \cap e_8 \cong so(16)$.

\subsection{Solvable Lie algebras.}

The problem of classifying solvable Lie algebras up to isomorphism
is, as far as we know, completely solved only for real Lie
algebras of dimension up to and including six (see for example
\cite{mor}--\cite{turk:}). The difficulty in the classification of
these algebras is, above all, connected with the fact that the
number of non-isomorphic Lie algebras increases considerably  with
increasing dimension, beginning with dimension five. Thus,
according to \cite{mub1:}, there are 66 classes of non-isomorphic
real, solvable Lie algebras of dimension five. Furthermore, for
dimension six, there are 99 classes of non-isomorphic algebras
just amongst the  real solvable algebras containing a nilpotent
element \cite{mub2:}.

Let us consider in more detail at the structure of solvable Lie
algebras over the field ${\bf R}$ with dimension no greater than
five. We give a method of searching for their realizations in the
class of differential operators.

Let $L_n$ denote a solvable Lie algebra of dimension $n,$ over a
field of characteristic zero. It is known (\cite{hel}) that there
exists a series of subalgebras

\[
L_n\supset L_2\supset \ldots \supset L_1\supset L_0=\{0\}
\]
such that each subalgebra $L_i\,(i=1,\ldots,n-1)$ is an ideal of
the algebra $L_{i+1}.$ This series is called the {\it composition
series} of the algebra $L_n.$

The existence of the composition series for a real solvable Lie
algebra allows us to make the following important conclusion: if,
in the given class of differential operators, there is a
realization of the solvable Lie algebras with $\dim L\leq m,$ and
there is no realization for algebras with $\dim L= m+1,$ then
those realizations which appear give a complete description of the
realizations of solvable algebras in the given class of vector
fields.

Further, we shall use the following notation: $A_{k.i}=\langle
e_1,\ldots, e_k\rangle$ denotes a Lie algebra of dimension $k,$
$e_j\,(j=1,\ldots,k)$ is its basis, and the index $i$ denotes the
number of the class to which the given Lie algebra belongs.

Fixing the type of the algebra $A_{k.i},$ we shall give only the
non-zero commutation relations between the basis elements. Among
the solvable Lie algebras over ${\bf R}$ of lowest dimension, we
have only one algebra which is one-dimensional $A_1=\langle
e_1\rangle,$ and two algebras which have dimension two:

\begin{eqnarray*}
A_{2.1}&=& \langle e_1, e_2 \rangle = A_1 \oplus A_1 = 2 A_1; \\
A_{2.2}&=& \langle e_1, e_2 \rangle , \ \ \ [e_1, e_2 ]= e_2.
\end{eqnarray*}

Further, we shall call {\it decomposable} Lie algebras those
algebras which can be decomposed as a direct sum of solvable
algebras of lower dimension. We give a list of all solvable Lie
algebras, up to and including dimension five, in Appendix 1.

\bigskip
It is clear that the search for realizations of solvable Lie
algebras over ${\bf R}$ must be begun with the description of the
inequivalent forms of the general infinitesimal operator, up to
equivalence under the transformations of $\cal E.$ Each of the
operators obtained will be a basis for the inequivalent
realizations of one-dimensional Lie algebras. Further, the
completion of the basis operators of each of the one-dimensional
Lie algebras, by an infinitesimal operator of the most general
form, is done by extension of the realizations of the
one-dimensional Lie algebras to realizations of two-dimensional
Lie algebras. In doing this, in order to simplify the form of the
second basis operator one uses those transformations from $\cal E$
which leave invariant the form of the first basis operator.

Analogously, the realizations of the two-dimensional Lie algebras
which one obtains, are extended to realizations of
three-dimensional solvable Lie algebras, and then the realizations
of the three-dimensional Lie algebras are extended in the same way
to realizations of the four-dimensional algebras, and so on.

In the extension of the realizations of Lie algebras of lower
dimension to realizations of decomposable solvable Lie algebras is
done simply by adding to each realization a basis operator which
commutes with all the other basis elements.

For the construction of the realizations of non-decomposable
solvable Lie algebras, as is shown by an analysis of their
structure above, one may also carry out the extension of the
realizations of Lie algebras of lower dimension to realizations of
non-decomposable solvable Lie algebras of higher dimension.

\subsection{Semi-direct sums of semi-simple and solvable Lie algebras.}

Lie algebras which are semi-direct sums of  semi-simple and
solvable Lie algebras can be divided into two classes:

\begin{itemize}

\item[1)] those Lie algebras which are direct sums of semi-simple and
solvable Lie algebras (decomposable algebras);

\item[2)] algebras which cannot be written as a direct sum of
semi-simple and solvable Lie algebras (indecomposable algebras).

\end{itemize}

Since decomposable algebras have the structure
\[
L=S\oplus N
\]
where $S$ is the Levi factor and $N$ is the radical (maximal
solvable ideal of $L$), then a complete description of these
algebras is easily obtained by combining the semi-simple and
solvable Lie algebras. However, since the classification of
solvable Lie algebras has only been done partially, there is a
corresponding incompleteness in the classification of decomposable
Lie algebras. The classification of indecomposable Lie algebras
has been done only as far as for Lie algebras of dimension eight
(\cite{turk1:}). These are Lie algebras whose Levi factor is
$sl(2,{\bf R})$ or $so(3).$

We give a complete list of these algebras in Appendix 2. We use the
following notation:
\begin{eqnarray*}
sl(2,{\bf R})=\langle e_1,e_2,e_3\rangle; \quad
[e_1,e_2]=2e_2,\,[e_1,e_3]=-2e_3,\, [e_2,e_3]=e_1\\ so(3)=\langle
e_1,e_2,e_3\rangle; \quad [e_1,e_2]=e_3,\,[e_1,e_3]=-e_2,\,
[e_2,e_3]=e_1.
\end{eqnarray*}

We note that the basis of $sl(2,{\bf R})$ given here differs from
that given previously, but it is not difficult to see that they
are isomorphic. Indeed, if we make the transformations
\[
e_1\to 2e_2,\;\, e_2\to e_3,\;\, e_3\to e_1
\]
then we have an isomorphism from the basis given here to the basis
given previously.

In denoting the radicals $N=\langle e_4,\ldots, e_m\rangle\,
(m=\dim\, N -3),$ we keep to the notation used for the
classification of solvable algebras given above. Moreover, the
corresponding commutation relations for the basis operators of $N$
can easily be obtained from those given by replacing the index $i$
by the index $i+3.$  Thus, in listing the algebras which are
semi-direct sums of semi-simple and solvable Lie algebras, we give
only those commutators $[e_i,e_j]=c^{k}_{ij}e_k,\, i=1,2,3;
j,k=4,\ldots,m$ which are non-zero.

Taking into account the above classification of finite-dimensional
real Lie algebras, we take a closer look at the second step of the
algorithm for the group classification of differential equations.

After having completed the first step of the algorithm, we have
the general form of the infinitesimal symmetry operator (together
with a defining system) for the given equation, and we have group
$\cal E$ of equivalence transformations of this equation.

At the beginning of the second step we have to bring the symmetry
operator to the simplest form, using transformations from $\cal
E.$ We note that it is well-known (\cite{o:}) that linearization
of the vector field is not possible since the group $\cal E$ is a
subgroup of all the local transformations of the manifold $V$ of
dependent and independent variables which enter into the
differential equation. Thus, we will obtain, up to equivalence,
some finite set of simplest forms for the symmetry operator.

Further, using the determining equations, we find from the
equation at hand, equations which admit the operators we have
obtained as symmetry operators. With this, we will obtain the
group classification of the differential equations of the given
form which admit one-dimensional Lie algebras of invariance.

The list of simplest forms of the symmetry operator which we find,
allows us to take one of the symmetry operators in its simplest
form, when we consider the realizations of Lie algebras of higher
dimension. Moreover, we may first consider realizations of
semi-simple and solvable Lie algebras.

When we consider realizations of semi-simple Lie algebras, we
must, as well as looking at those semi-simple Lie algebras which
appear in our list, also take into account semi-simple Lie
algebras which are direct sums of semi-simple Lie algebras which
do not appear in the list given above.

We take a closer look at low-dimensional semi-simple Lie algebras.
As we noted above, the semi-simple Lie algebras of lowest
dimension are the algebras $sl(2,{\bf R})$ and $so(3),$ both
having dimension 3. Then we have Lie algebras of dimension 6 (the
algebras $so(4) \cong so(3) \oplus so(3), so(3,1), so(2,2) \cong
sl(2,{\bf R}) \oplus sl(2,{\bf R}), so^{*}(4) \cong so(3) \oplus
sl(2,{\bf R})$); dimension 8 (the algebras $su(3), sl(3,{\bf R}),
su(2,1)$).These algebras can be found in our list. However, the
semi-simple Lie algebras of dimension 9 (the algebras $so(4)
\oplus so(3), so(4) \oplus sl(2,{\bf R}), so(3,1) \oplus so(3),
so(3,1) \oplus sl(3,{\bf R}),$ $so(2,2) \oplus so(3), so(2,2)
\oplus sl(2,{\bf R})$) are not to be found in our list.

For solvable Lie algebras, one may extend realizations of lower
dimensional algebras to realizations of higher-dimensional
algebras according to the scheme given above. Moreover, for
solvable Lie algebras of higher dimension, the composition series
play an important role: if there is a realization in terms of
operators of a given class of solvable Lie algebras of dimension
$m-1,$ but no realization of a solvable Lie algebra of dimension
$m,$ the there will, {\it a priori}, be no realizations of
solvable Lie algebras of dimension greater than $m.$

When looking at the realization of Lie algebras which are
semi-direct sums of semi-simple and solvable  Lie algebras, one
must extend a given realization of semi-simple Lie algebras by
operators which will be basis operators of the corresponding
radicals. Moreover, it is only necessary to take into account
those radicals which are isomorphic to solvable Lie algebras which
have realizations in the given class of vector fields.

Finally, we note that since the classification of solvable Lie
algebras, and those algebras which are semi-direct sums of
semi-simple and solvable Lie algebras, is incomplete, then it is
not possible to give a complete group classification of
differential equations within the present framework. However, the
problem of the complete group classification of, for instance,
scalar equations in two-dimensional space-time which are invariant
under finite-dimensional Lie algebras, is constructive in this
approach.

\section{Classification results}
\setcounter{equation}{0} \setcounter{thm}{0} \setcounter{rmk}{0}
\setcounter{lemma}{0}

First of all let us mention some papers in which group classification
of particular equations of the form (\ref{2.1}) has been
carried out.

\begin{tabular}{lll}
Ovsjannikov\ (1959) & $F=F(u)$, &
$G=\displaystyle{\frac{dF}{du}}u^{2}_{x}$\ \cite{ov1};\\ [4mm]
Akhatov\ et al\ (1987) & $F=F(u_{x})$,& $G=0$ \ \cite{ah:};\\
[3mm] Dorodnitsyn\ (1982) & $F=F(u)$, &
$G=\displaystyle{\frac{dF}{du}}u^{2}_{x}+g(u)$ \ \cite{dor};\\
[3mm] Oron \&\ Rosenau\ (1986), &  \\ Edwards\ (1994) & $F=F(u)$,
& $G=\displaystyle{\frac{dF}{du}}u^{2}_{x}+f(u)u_{x}$ \
\cite{oro,edw};
\\ [3mm] Gandarias\ (1996) & $F=u^{n}$, &
$G=\displaystyle{\frac{dF}{du}}u^{2}_{x}+g(x)u^{m}u_{x}
+f(x)u^{s}$ \ \cite{gan};
\\
[3mm] Cherniha\ \&\ Serov\ (1998) & $F=F(u)$, &
$G=\displaystyle{\frac{dF}{du}}u^{2}_{x}+f(u)u_{x}+g(u)$ \
\cite{che};
\\ [3mm] Zhdanov\ \&\ Lahno\ (1999) & $F=1$, & $G=G(t,x,u,u_x)$ \
\cite{zhd99}.
\end{tabular}

We shall apply the algorithm described above  in order to perform
an exhaustive group classification of invariant equations of the
general form (\ref{2.1}). That is, we shall describe all
inequivalent forms of functions $F, G$ such that the corresponding
equation admits a non-trivial symmetry group.

\subsection{Computation of the equivalence group admitted by equation
(\ref{2.1})}

The first step of the algorithm is the determination of the most
general form of the infinitesimal symmetry operator admitted by
the PDE (\ref{2.1}). To this end, we use Lie's method
\cite{lie24}--\cite{ol1:} and look for a symmetry generator in the
form
\begin{eqnarray}
Q=\tau (t,x,u) \,\partial_t + \xi (t,x,u) \,\partial_x +
\eta(t,x,u) \,\partial_u, \label{n2}
\end{eqnarray}
where\ $\tau $,\ $\xi $,\ $\eta $ are arbitrary, real-valued
smooth functions defined in some subspace of the space\
$V=X\otimes R^1$\ of the independent variables\ $X=\langle t,x
\rangle $\ and the dependent variable\ $R^1=\langle u \rangle$.

As a result, we find that the operator (\ref{n2}) generates a one-parameter
symmetry
group of equation (\ref{2.1}) iff
\begin{equation}
\varphi ^t-[\tau F_t+\xi F_x+\eta F_u+\varphi ^xF_{u_{x}}]u_{xx}-
\varphi ^{xx}F-\tau G_t-\xi G_x-\eta G_u-\varphi ^xG_{u_{x}}
\Big|_{u_{t}=Fu_{xx}+G}=0, \label{n3}
\end{equation}
where
\begin{eqnarray}
\varphi ^t  &=& D_t(\eta )-u_tD_t(\tau )-u_xD_t(\xi ), \nonumber
\\ \varphi ^x  &=& D_x(\eta )-u_tD_x(\tau )-u_xD_x(\xi ),
\nonumber \\ \varphi ^{xx}&=& D_x(\varphi ^x)-u_{tx}D_x(\tau
)-u_{xx}D_x(\xi ) \nonumber
\end{eqnarray}
and\ $D_t$,\ $D_x$ are operators of total differentiation in $t$
and $x$ respectively.

After simplifying (\ref{n3}) we arrive at the following assertion.
\begin{lemma}
The symmetry group of the nonlinear heat equation PDE (\ref{2.1})
is generated by the infinitesimal operators of the form
\begin{equation}
Q=a(t)\,\partial_t+b(t,x,u)\,\partial_x+c(t,x,u)\,\partial_u,
\label{n4}
\end{equation}
where\  $a,\ b, \ c$\  are real-valued functions that satisfy the
system of PDEs
\begin{eqnarray}
 && (2b_x+2u_xb_u-\dot a)F=aF_t+bF_x+cF_u+(c_x+u_xc_u-u_xb_x-u^2_xb_u)
F_{u_{x}}, \nonumber \\ && c_t-u_xb_t+(c_u-\dot
a-u_xb_u)G+(u_xb_{xx}-c_{xx}-2u_xc_{ux}- u^2_{x}c_{uu}+ \label{n5}
\\
 && + 2u^2_xb_{xu}+u^3_xb_{uu})F
  = aG_t+bG_x+cG_u+(c_x+u_xc_u-u_xb_x-u^2_xb_u)G_{u_{x}}.\nonumber
\end{eqnarray}
\end{lemma}
In the rest of this paper we use the notation\ $\dot
a=\displaystyle{\frac{da}{dt}}$, \ $\ddot
a=\displaystyle{\frac{d^2a}{dt^2}}$.

In order to construct the equivalence group\ $\cal E$\ of the
class of PDEs (\ref{2.1}) one has to select from the set of
invertible changes of variables of the space\  $V$
\begin{eqnarray}
\bar t=\alpha (t,x,u),\ \ \bar x=\beta (t,x,u),\ \  v =\gamma
(t,x,u),\ \ \displaystyle{\frac{D(\alpha, \beta, \gamma
)}{D(t,x,u)}}\not =0,\label{n6}
\end{eqnarray}
those changes of variable which do not alter the form of the class of PDEs
(\ref{2.1}).
\begin{lemma}
The maximal equivalence group\ $\cal E$\  of the class of PDEs
(\ref{2.1}) reads as
\begin{eqnarray}
\bar t=T(t), \ \ \bar x=X(t,x,u), \ \  v =U(t,x,u), \label{n7}
\end{eqnarray}
where\ $\dot T \not =0,$ \ $\displaystyle{\frac{D(X,U)}
{D(x,u)}}\not =0$.
\end{lemma}
{\bf Proof.}\ Let (\ref{n6}) be an invertible change of
variables that transforms equation (\ref{2.1}) into another
equation of the same form (\ref{2.1}), namely,
\begin{eqnarray}
v_{\bar  t}=\tilde F(\bar t,\bar x,v ,v_{\bar x})v_{\bar x \bar x}
+\tilde G(\bar t,\bar x,\bar v ,\bar v_x). \label{n8}
\end{eqnarray}
Computing $u_x$ according to (\ref{n6}) we get
\begin{eqnarray}
u_x=\displaystyle{\frac{v_{\bar t}\alpha_x+v_{\bar x}\beta_x
-\gamma_x}{\gamma_u-v_{\bar t}\alpha_u-v_{\bar x}\beta_u}}.
\label{n9}
\end{eqnarray}
As the functions \ $F$, \ $G$ \  in (\ref{2.1}) and\ $\tilde F$, \
$\tilde G$ \ in (\ref{n8}) are arbitrary functions of the
corresponding arguments, we must have that $$
u_x \rightarrow g(\bar t, \bar x,v ,v_{\bar x}) $$ for some function $g.$
This implies that\
$\alpha_x=\alpha_u=0$ in (\ref{n9}). Consequently,\ $\alpha
=T(t),\ \dot T \not =0.$

Next, making the change of variables (\ref{n6}), where\ $\alpha =
T(t)$, we arrive at the relations
\begin{eqnarray}
&u_t \rightarrow v_{\bar t}\dot T(\gamma_u- v_{\bar x}
\beta_u)^{-1} + \theta_1(\bar t, \bar x, v, v_{\bar x}), \nonumber
\\ &u_{xx} \rightarrow v_{\bar x \bar x}\theta_2(\bar t,\bar x, v,
v_{\bar x}) + \theta_3(\bar t,\bar x,v ,v_{\bar x}), \label{n10}
\end{eqnarray}
where\ $\theta_1, \ \theta_2\ne 0, \ \theta_3$\ are some functions
of $\alpha,\ \beta,\ \gamma$ and of their derivatives. Then, inserting\
$u_t, \ u_{xx}$ from (\ref{n10})\ into
(\ref{2.1}), we arrive at a PDE of the form (\ref{n8}). The lemma is
proved.

\subsection{Classification of equations (\ref{2.1}) invariant under
semi-simple Lie algebras}

Now we proceed to solving the classification problem for the
nonlinear heat-conductivity equation (\ref{2.1}). Our first step
is to construct realizations of finite-dimensional real Lie
algebras whose representation space is spanned by operators of the
form (\ref{n4}). It should be noted that the realizations are
constructed up to equivalence as determined by the transformations
(\ref{n7}). In the second step, we choose those realizations which
are invariance algebras of PDE (\ref{2.1}) and thus specify the
form of the functions\ $F, G$. Finally, in the third step, we find
the maximal symmetry groups of the equations we obtain, and thus
complete the group classification of PDE (\ref{2.1}).

By the Levi-Mal'cev theorem, we need only consider the cases of
semi-simple, solvable and semi-direct sums of semi-simple and
solvable symmetry algebras. This will yield an exhaustive description of
invariant equations of the form (\ref{2.1}). In this subsection, we analyze
the case of
semi-simple symmetry algebras. As semi-simple Lie algebras can
always be decomposed into a direct sum of simple Lie algebras (which is the
content of Cartan's theorem),
we begin with the lowest dimensional simple Lie algebras\ $sl(2,{\bf R})$\
and\ $so(3)$.

We begin by proving the following useful lemma:
\begin{lemma}
There are changes of variables (\ref{n7}), that reduce an operator
(\ref{n4}) to one of the operators below:
\begin{eqnarray}
 Q &=& \partial_t, \label{n11} \\
 Q &=& \partial_x. \label{n12}
\end{eqnarray}
\end{lemma}
{\bf Proof.}\ Making the change of variables (\ref{n7}) transforms
operator (\ref{n4}) to the following one:
\begin{eqnarray}
Q\rightarrow Q' =a\dot T \,\partial_ {\bar t}+(aX_t+bX_x+cX_u)
\,\partial_{\bar x}+(aU_t+bU_x+cU_u)\,\partial_v. \label{n13}
\end{eqnarray}

Suppose\ $a\not =0.$ Then, choosing in (\ref{n7})
the function\ $T$ \ to be a solution of the equation\ $\dot
T=a^{-1}$ and the functions\ $X$\ and\ $U$\ to be independent fundamental
solutions of the  first-order PDE $$
aY_t+bY_x+cY_u=0, \ \ Y=Y(t,x,u), $$ we find that the operator
(\ref{n13}) takes the form $Q'=\partial_{\bar t}$.

Now suppose\ $a=0$. Then\ $b^2+c^2 \not =0$. If\ $b\not
=0$, then choosing in (\ref{n7}) a particular solution of PDE
$bX_x+cX_u=1$ as the function\ $X$\ and a fundamental solution of
PDE $bU_x+cU_u=0$ as the function\ $U$,\ we transform (\ref{n13}) to
become $Q'=\partial_{\bar x}$.

If\ $b=0$,\ $c\not =0$, then making the change of variables (\ref{n7})
with $\bar t=t,\ \bar x=u,\ v=x,$ we again get the case
$b\not = 0$.

By the direct calculation we can verify that there is no
transformation from\ $\cal E$,\ that reduce operator (\ref{n11})
to the form (\ref{n12}).

The lemma is proved.

\begin{thm}
Within the equivalence relation\ $\cal E,$\ there exists only one
realization of the algebra\ $so(3)$\ by operators of the form (\ref{n4}):
\begin{eqnarray}
\langle \partial_x,\ \tan u\, \sin x\, \partial_x + \cos x\,
\partial_u,\ \tan u\, \cos x\, \partial_x - \sin x\, \partial_u
\rangle , \label{n14}
\end{eqnarray}
It is the invariance algebra of an equation from the class
(\ref{2.1}). Furthermore, the most general form of the functions\
$F, G$ allowing for PDE (\ref{2.1}) to be invariant under the
above realization is given by
\begin{equation}
F=\displaystyle{\frac{\sec ^2 u}{1+\omega ^2}},\quad
G=\displaystyle{\frac{2\omega ^2+1}{1+\omega ^2}}\tan u +
\sqrt{1+\omega ^2}\tilde G(t),\quad \omega =u_x\sec u. \label{n15}
\end{equation}
Provided\ the function $\tilde G$ is arbitrary, the realization
(\ref{n14}) is the maximal symmetry algebra of the corresponding
equation (\ref{2.1}).
\end{thm}
{\bf Proof.}\ The Lie algebra\ $so(3)=\langle Q_1, Q_2, Q_3
\rangle $\ is defined by the following commutation relations:
\begin{eqnarray}
\lbrack Q_1, Q_2 \rbrack =Q_3,\quad \lbrack Q_1, Q_3 \rbrack
=-Q_2, \quad \lbrack Q_2, Q_3 \rbrack =Q_1. \label{n16}
\end{eqnarray}
To describe all inequivalent realizations of the algebra\ $so(3)$\
we take operators of the form (\ref{n4}) as the basis elements\
$Q_i$\ $(i=1,2,3)$\ of\ $so(3)$\ and then study the restrictions
imposed on their coefficients by relations (\ref{n16}). We also use
transformations (\ref{n7}) in order to simplify the
final forms of the basis elements.

In view of Lemma 3.3, we can take one of the basis elements of the
algebra\ $so(3)$\ (say,\ $Q_1$)\ either in the form\ $\partial_t$\
or\ $\partial_x$.

Let\ $Q_1=\partial_t$. Using the first two
commutation relations from (\ref{n16}) yields
\begin{eqnarray}
Q_2 &=& \lambda \cos t\, \partial_t + \lbrack b\cos t + \beta \sin
t \rbrack\, \partial_x + \lbrack c \cos t + \gamma \sin t
\rbrack\,
\partial_u, \nonumber \\ Q_3 &=& -\lambda \sin t\, \partial_t + \lbrack
- b\sin t + \beta \cos t \rbrack\, \partial_x + \lbrack - c\sin t
+ \gamma \cos t \rbrack\, \partial_u, \nonumber
\end{eqnarray}
where\ $\lambda = {\rm const}\ \in {\bf R},\  b=b(x,u),\
c=c(x,u),\ \beta = \beta (x,u),\ \gamma = \gamma (x,u)$ are
arbitrary smooth functions.Then, using the third commutation
relation, we arrive at the equation\ $\lambda ^2=-1$\ which has no
real solutions $\lambda$. Consequently, in the case when the
operator\ $Q_1$\ is equivalent to the operator\ $\partial_t,$ there
are no realizations of the algebra\ $so(3)$.

Turn now to the case\ $Q_1=\partial_x$. As a straightforward calculation
shows, the
most extensive subgroup of the equivalence group\ $\cal E$\ not
altering the form of\ $Q_1$ is of the form
\begin{eqnarray}
\bar t=T(t),\ \bar x=x+X(t,u),\ v=U(t,u),\ \dot T\not =0,\ U_u\not
=0. \label{n17}
\end{eqnarray}

Using the first two commutation relations from
(\ref{n16}) we get
\begin{eqnarray}
Q_2 &=& \alpha \cos (x+\gamma )\,\partial_x + \beta \cos (x+\theta
)\,
\partial_u, \nonumber \\
Q_3 &=& -\alpha \sin (x+\gamma )\,\partial_x - \beta \sin
(x+\theta )\,
\partial_u, \label{n18}
\end{eqnarray}
where\ $\alpha = \alpha (t,u),\ \gamma = \gamma (t,u),\ \beta =
\beta (t,u),\ \theta = \theta (t,u)$\ are arbitrary smooth
functions. Now, either\ $\beta =0$ \ or \ $\beta \not =0.$ If \ $\beta =
0,$ the third commutation
relation givews\ $\alpha ^2=-1,$ which has no real
solutions. Consequently,\ $\beta \not =0$.

Choosing in (\ref{n17})\ $X=\theta$\ and furthermore, taking an
arbitrary solution of the equation\ $U_u=\beta ^{-1}$,\ as\ $U$,\
we simplify the forms of the operators\ $Q_2, \ Q_3$ to obtain
\begin{eqnarray*}
Q_2 &=& \alpha \cos (x+\gamma )\,\partial_x + \cos x\,\partial_u,
\\ Q_3 &=& -\alpha \sin (x+\gamma )\,\partial_x - \sin
x\,\partial_u,
\end{eqnarray*}
where\ $\alpha =\alpha (t,u),\ \gamma =\gamma (t,u)$\ are
arbitrary smooth functions (here and in the following,  we keep the initial
designations for the transformed operators to simplify the notation).

The third commutation relation for the operators \ $Q_1,\ Q_2$\  which we
have obtained,
yields the equations\ $\cos \gamma =0,\ \alpha ^2 +
\alpha_u \sin \gamma =-1$, whence
\begin{eqnarray*}
Q_2 &=& \tan \lbrack u\pm \tilde \alpha (t)\rbrack \sin x\,
\partial_x + \cos x\,\partial_u,\\ Q_3 &=& \tan \lbrack u\pm
\tilde \alpha (t)\rbrack \cos x\, \partial_x - \sin x\,\partial_u,
\end{eqnarray*}
where\ $\tilde \alpha (t)$\ is an arbitrary smooth function.

Finally, putting\ $T=t,\ X=0,\ U=u\pm \tilde \alpha (t)$\  in
(\ref{n17}) we find that the above realization is equivalent
to (\ref{n14}).

To complete the proof we have to verify whether there exists an
equation of the form (\ref{2.1}), whose symmetry algebra contains
subalgebra (\ref{n14}). Invariance of (\ref{2.1}) with respect to
the one-parameter group having the generator\ $Q_1$\ means that\
$F=F(t,u,u_x),\ G=G(t,u,u_x)$. Writing down condition (\ref{n5})
for the operators\ $Q_2$,\ $Q_3$\ we get the following system of
PDEs:
\begin{eqnarray*}
&&F_u-u_x\tan u F_{u_{x}} = 2 \tan u F,\\ &&(1 + u^2_x \sec ^2
u)F_{u_{x}} = -2u_x \sec ^2 u F, \\ && u_x \sec ^2 u G + u_x \tan
u(1 - 2u^2_x\sec ^2 u)F =(1 + u^2_x\sec ^2u) G_{u_{x}},\\
&&(1+2u^2_x\sec ^2u)F = G_u-u_x \tan u G_{u_{x}}.
\end{eqnarray*}

Solving the first two equations of the above system gives
\[
F=\displaystyle{\frac{\sec ^2u}{1+\omega ^2}}\tilde F(t),
\]
where\ $\omega =u_x\sec u$. Integrating  the fourth equation
yields \ $G=\displaystyle{\frac{2\omega ^2+1}{1+\omega ^2}}
\tan u \tilde F(t) + \bar G(t,\omega )$. Finally, solving the
third equation we get the form of\ $\bar G(t, \omega)$
\[
\bar G(t, \omega)=\sqrt{1+\omega ^2}\tilde G(t).
\]
Thus PDE (\ref{2.1}) is invariant with respect to the algebra
(\ref{n14}) iff
\begin{equation}
F=\displaystyle{\frac{\sec^2 u}{1+\omega ^2}}\tilde F(t), \
G=\displaystyle{\frac{2\omega ^2+1}{1+\omega ^2}}\tan u\tilde F(t)
+ \sqrt{1+\omega ^2}\tilde G(t), \ \omega =u_x\sec u. \label{n19}
\end{equation}
with arbitrary smooth functions\ $\tilde F(t)$, \ $\tilde G(t)$
provided that\ $\tilde F(t) \not =0$.

Evidently, the change of variables (\ref{n17}) with\ $X=0,\ U=u$\
does not alter the forms of the operators of the realization
(\ref{n14}). Choosing a solution of the equation\ $\dot T=\tilde
F$\ as\ $T$,\ we get\ $\tilde F(t)=1$. By direct computation
one shows that if the function\ $\tilde G(t)$\ is arbitrary, then the
algebra (\ref{n14}) is the maximal invariance algebra admitted by
the equation obtained. The theorem is proved.

\begin{thm}
There exist five inequivalent realizations of the algebra\
$sl(2,{\bf R})$ by operators (\ref{n4}), which are admitted by
PDEs of the form (\ref{2.1})
\begin{eqnarray}
\langle 2t\,\partial_t+x\,\partial_x,\
-t^2\,\partial_t-tx\,\partial_x +x^2\,\partial_u,\ \partial_t
\rangle, \label{n20}\\ \langle 2t\,\partial_t+x\,\partial_x,\
-t^2\,\partial_t+x(x^2-t)\,\partial_x,\
\partial_t \rangle, \label{n21}\\
\langle 2x\,\partial_x - u\,\partial_u, -x^2\,\partial_x +
xu\,\partial_u,\
\partial_x \rangle, \label{n22}\\
\langle 2x\,\partial_x-u\,\partial_u,\
(u^{-4}-x^2)\,\partial_x+xu\,\partial_u,\
\partial_x \rangle, \label{n23}\\
\langle 2x\,\partial_x-u\,\partial_u,\
-(u^{-4}+x^2)\,\partial_x+xu\,\partial_u,\
\partial_x \rangle. \label{n24}
\end{eqnarray}
The forms of the functions\ $F, G$\ determining the corresponding
invariant equations are given as follows:\vspace{3mm}
\begin{center}
\begin{tabular}{|c|c|c|}\hline
$sl(2,{\bf R})$       & $F$ & $G$ \\ \hline & & \\(\ref{n20}) &
$\tilde F(\omega )$ & $x^{-2}\left[ \tilde G(\omega )-2u \tilde
F(\omega )+u^2-u\omega \right], \ \omega =2u-xu_x$ \\ & & \\
(\ref{n21}) & $\omega ^{-3}$ & $x^{-2}\left[
-\displaystyle{\frac{1}{4}} \omega +3\omega ^{-2}+\omega
^{-1}\tilde G(u)\right], \ \omega =xu_x$
\\ & & \\
(\ref{n22}) & $u^{-4}$ & $-2u^{-5}u_{x}^2$\\ & & \\ (\ref{n23}) &
$u^{-4}\left( 1+4\omega ^2 \right) ^{-1} $ & $u\left[
\sqrt{1+4\omega ^2}\tilde G(t)- \displaystyle{\frac{10\omega
^2+1}{8\omega ^2+2}}\right] , \ \omega =u^{-3}u_x$ \\ & & \\
(\ref{n24}) & $u^{-4}\left( 1-4\omega ^2 \right) ^{-1}$ & $u\left[
\sqrt{|1-4\omega ^2|}\tilde G(t)+ \displaystyle{\frac{10\omega
^2-1}{8\omega ^2-2}}\right], \ \omega =u^{-3}u_x$ \\ & & \\ \hline
\end{tabular}
\end{center}
\vspace{3mm}

\noindent If the functions\ $\tilde F,\ \tilde G$ are arbitrary,
then the corresponding realizations of the algebra\ $sl(2,{\bf
R})$\ are maximal invariance algebras of the respective PDEs
(\ref{2.1}). Furthermore, the maximal symmetry group admitted by
the third PDE from the above list
\[
u_t=u^{-4}u_{xx} - 2u^{-5}u^2_x
\]
is the five-dimensional Lie algebra\ $sl(2,{\bf R})\oplus
L_{2.1}$, where\ $sl(2,{\bf R})$\ is given in (\ref{n22}) and\
$L_{2.1}=\langle 4t \,\partial_t+u\,\partial_u,\ \partial_t
\rangle $.
\end{thm}
{\bf Proof.}\ The Lie algebra\ $sl(2,{\bf R})=\langle Q_1,\ Q_2,\
Q_3 \rangle $\ is defined by the following commutation relations:
\begin{equation}
[Q_1,\ Q_2]=2Q_2, \quad [Q_1,\ Q_3]=-2Q_3, \quad [Q_2,\ Q_3]=Q_1.
\label{n25}
\end{equation}
In view of Lemma 3.3 we can choose the operator\ $Q_3$\ either in
the form $\partial_t$\ or\ $\partial_x$.

Let\ $Q_3=\partial_t$. Imposing the second commutation relation
from (\ref{n25}) gives (up to  equivalence under\
${\cal E}$)\ the operator\ $Q_1$\ either equals to\
$2t\,\partial_t$\ or \ $2t\,\partial_t+x\,\partial_x$.

If\ $Q_1=2t\,\partial_t$,\ then it follows from the remaining
commutation relation that\ $Q_2=-t^2\,\partial_t$,\ so that we obtain
the realization\ $\langle 2t\,\partial_t,\ -t^2\,\partial_t,\
\partial_t \rangle$. However, PDE (\ref{2.1}) can admit this
algebra only when the condition\ $F=0$ holds. This contradicts the
assumption\ $F\not =0$ and, consequently, there are no
corresponding invariant equations within the class (\ref{2.1}).

If now\ $Q_1=2t\,\partial_t + x\,\partial_x$,\ we get
(up to  equivalence under\ ${\cal E}$)\  the realization\
$\langle 2t\,\partial_t + x\,\partial_x,\ -t^2\,\partial_t -
tx\,\partial_x,\
\partial_t \rangle$\ and the realizations (\ref{n20}), (\ref{n21}) of the
algebra\ $sl(2,{\bf R})$. Substituting into the  invariance conditions
(\ref{n5}) shows that the first realization cannot be admitted by
PDE (\ref{2.1}). This leaves us with the realizations (\ref{n20}),
(\ref{n21}).

The most general equations (\ref{2.1})
invariant with respect to realizations (\ref{n20}), (\ref{n21})
are given by:
\begin{eqnarray*}
&&F=\tilde F(\omega ),\quad G=x^{-2}\left[ \tilde G(\omega )- 2u
\tilde F(\omega ) + u^2 - u\omega \right],\quad \omega =2u-xu_x,\\
&&F=\omega ^{-3}\tilde F(u),\quad G=x^{-2}\left[
-\displaystyle{\frac{1}{4}}\omega + 3\omega ^{-2}\tilde F(u)+
\omega ^{-1}\tilde G(u)\right],\quad \omega =xu_x.
\end{eqnarray*}
It is not difficult to show that the change of variables\
\[
t=t, \quad x=x, \quad u=U(v), \quad U'\not =0, \quad v=v(t,x)
\]
does not alter the form of the basis operators of the realization
(\ref{n21}). So, choosing the function\ $U$\ to be a solution of the
equation\ $(U')^3=\tilde F(U)$, we can
transform PDE (\ref{2.1}) invariant under (\ref{n21}) in such a
way that\ $\tilde F\equiv 1$.

We now turn to the case\ $Q_3=\partial_x$. Using the commutation
relations (\ref{n25}) we find that the inequivalent realizations of\
$sl(2,{\bf R})$ within the class of operators (\ref{n4}) are
exhausted by the realization\ $\langle 2x\,\partial_x,\
-x^2\,\partial_x,\ \partial_x\rangle $\ and by the realizations
(\ref{n22}), (\ref{n23}), (\ref{n24}). The invariance
conditions (\ref{n5}) show that the first realization cannot be
an invariance algebra of PDE of the form (\ref{2.1}).  The
remaining realizations are invariance algebras of PDEs (\ref{2.1})
under proper specification of the functions\ $F, G:$
\begin{eqnarray}
&&F=u^{-4}\tilde F(t),\quad G=-2u^{-5}u_{x}^2\tilde F(t)+u\tilde
G(t),\ (\mbox{ for the realization\ (\ref{n22})}); \label{temp1}\\
&&F=\displaystyle{\frac{1}{u^4\left( 1+4\omega ^2 \right) }}\tilde
F(t), \quad G=u\left[ \sqrt{1+4\omega ^2}\tilde G(t)-
\displaystyle{\frac{10\omega ^2+1}{8\omega ^2+2}}\tilde F(t)
\right], \nonumber \\ &&\quad \omega =u^{-3}u_x,\ (\mbox{
for the realization\ (\ref{n23}}); \label{temp2}\\
&&F=\displaystyle{\frac{1}{u^4\left( 1-4\omega ^2 \right) }}\tilde
F(t), \quad G=u\left[ \sqrt{|1-4\omega ^2|}\tilde G(t)+
\displaystyle{\frac{10\omega ^2-1}{8\omega ^2-2}}\tilde F(t)
\right], \nonumber\\ &&\quad \omega =u^{-3}u_x,\ (\mbox{
for the realization\ (\ref{n24})}).\label{temp3}
\end{eqnarray}

As the change of variables
\[
\bar t=T,\quad \bar x=x,\quad v=U(t)u,\quad T\not =0,\quad U\not
=0
\]
does not alter the form of the basis operators of the realization
(\ref{n22}), we can use it in order to simplify the forms of\ $F,
G$. Choosing the functions\ $T$\ and\ $U$\ to be
solutions of the equations \ $\dot U=U\tilde G(t)$,\ $U\not =0,$\
$\dot T=\tilde FU^4,$ we obtain\ $\tilde F\equiv 1,$\ $\tilde G\equiv
0$ in (\ref{temp1}).

Similarly, using the change of variables
\[
\bar t=T(t),\quad \bar x=x,\quad v=u
\]
which preserve the form of the basis operators of the realizations
(\ref{n23}), (\ref{n24}) we can choose\ $\tilde F\equiv 1$ in
(\ref{temp2}), (\ref{temp3}).

Computing the maximal invariance algebra of the PDE which admits the
realization (\ref{n22}) we get the five-dimensional Lie algebra
which is the direct sum of\ $sl(2,{\bf R})$ \ having the basis
elements (\ref{n22}) and the\ two-dimensional solvable Lie algebra\
$L_{2.1} = \langle 4t\,\partial_t + u\,\partial_u,\ \partial_t
\rangle.$

The remaining invariant equations contain an arbitrary function.
If there are no additional constraints on this function, then the
realizations (\ref{n20}), (\ref{n21}), (\ref{n23}), (\ref{n24})
of\ $sl(2,{\bf R})$\ are easily shown to be the maximal invariance
algebras of the corresponding invariant equations.

The theorem is proved.

\begin{thm}
The realizations of the algebras\ $so(3)$\ and\ $sl(2,{\bf R})$,
given in Theorems 3.1, 3.2, exhaust the set of all possible
realizations of semi-simple Lie algebras by operators (\ref{n4})
which are admitted by PDEs of the form (\ref{2.1}).
\end{thm}
{\bf Proof.}\ The simple Lie algebras of the lowest dimension
admit the following isomorphisms:
\[
su(2) \sim so(3) \sim sp(1), \ sl(2,{\bf R}) \sim su(1,1) \sim
so(2,1) \sim sp(1,{\bf R}).
\]
>From this it follows that the realizations given in Theorems 3.1,
3.2 exhaust the set of all possible realizations of
three-dimensional simple Lie algebras which are symmetry algebras
of (\ref{2.1}).

The next admissible dimension for simple Lie algebras is six.
There are four distinct six-dimensional simple Lie algebras over
the field of real numbers, namely,\ $so(4),\ so(3,1),\ so(2,2),$\
and\ $so^*(4)$.

As\ $so(4)=so(3)\oplus so(3)$,\ we have\ $so(4)=\langle Q_i,\, K_i
|\ i=1, 2, 3 \rangle$, where\ $\langle Q_1,\, Q_2,\, Q_3 \rangle =
so(3), \ \langle K_1, K_2, K_3 \rangle =so(3)$,\ and we have the
commutation relations,\ $[Q_i,\, K_j]=0, \ i, j=1, 2, 3$. Making
use of Theorem 3.1 we put the basis operators\ $Q_i\ (i=1, 2, 3)$\ to be
equal to the
corresponding basis operators of realization (\ref{n14}). Next,
the commutation relations\ $[Q_i,\, K_j]=0,\ (i, j=1, 2, 3)$\
imply the following form of the operators\ $K_j$:
\begin{eqnarray}
K_j=a_j(t)\,\partial_t,\quad a_j\not = 0,\ \ j=1,2,3. \label{n26}
\end{eqnarray}

Using the change of variables
\[
\bar t=T(t),\quad  \bar x = x,\quad v=u,
\]
(which does not change the form of operators (\ref{n14})), we can
transform the operator\ $K_1$\ to become\ $K_1=\partial_t$.
Checking the commutation relations for the algebra\ $so(3)$\
yields that \[ K_2=\lambda \cos (t+\lambda_1)\,\partial_t,\quad
K_3=-\lambda \sin (t+\lambda_1),
\]
where\ $\{\lambda_1, \lambda\} \subset {\bf R}$\ with\ $\lambda
^2=-1$. Consequently, there are no realizations of the algebra\
$so(4)$\ within the class of operators (\ref{n4}), which are
symmetry algebras of (\ref{2.1}).

We have the relation\ $so^{*}(4) \sim so(3)\oplus sl(2,{\bf
R})$. So,  in order to construct realizations of\ $so^{*}(4)$\ we have
to describe realizations of the algebra\  $sl(2,{\bf R})$\ by
operators of the form (\ref{n26}). Now, in proving Theorem
3.2, we  established, in particular, that there is a unique
realization of\ $sl(2,{\bf R})$\ by operators (\ref{n26}),
$\langle 2t\,\partial_t,\ -t^2\,\partial_ t,\
\partial_t \rangle $, which, however, cannot be admitted by a PDE of the form
(\ref{2.1}). This eliminates \ $so^{*}(4)$.

The algebra\ $so(3,1)$\ admits the Cartan decomposition\ $\langle
Q_1,\, Q_2,\, Q_3 \rangle \dot + \langle N_1,\, N_2,\, N_3
\rangle$, where\ $\langle Q_1,\, Q_2,\, Q_3 \rangle = so(3), \
[Q_i,\ N_j]=N_{k},\ [N_i,\ N_j]=-Q_{k},$\ $i,j,k={\rm
cycle}(1,2,3)$. Thus, taking as\ $Q_i \ (i=1, 2, 3)$\ the
corresponding basis operators of realization (\ref{n14}) and
computing the forms of the operators\ $N_1, N_2, N_3,$\ we get
within the equivalence relation ${\cal E}$ the following
relations:
\[
N_1 = \cos u \,\partial_u,\quad N_2 = -\sec u \ \cos x
\,\partial_x + \sin u \ \sin x \,\partial_u,\quad N_3 = \sec u \
\sin x \,\partial_x + \sin u \ \cos x \,\partial_u.
\]
Imposing the invariance conditions (\ref{n5}) for the operator\
$N_1$\ gives that\ $F=0$,\ contradicting  our initial
assumption\ $F\not =0$.

In studying realizations of the algebra\ $so(2,2)$\ we use the
fact that\ $so(2,2) \sim sl(2,{\bf R}) \oplus sl(2,{\bf R})$. In
view of this, we can choose the basis operators of this algebra so
that\ $so(2,2)= \langle Q_i,\, K_i\ |\ i=1, 2, 3 \rangle$,\ where\
$\langle Q_1,\, Q_2,\, Q_3 \rangle = sl(2,{\bf R}),\ \langle
K_1,\, K_2,\, K_3 \rangle = sl(2,{\bf R})$,\ and ,\
$[Q_i\ K_j]=0,\ i, j=1, 2, 3$. Now we can take as $Q_1, Q_2, Q_3$
the corresponding basis operators of the realizations of\ $sl(2,{\bf
R})$\ given by (\ref{n20})--(\ref{n24}). However, as  further
analysis shows, these realizations cannot be extended  to a
realization of\ $so(2,2)$ which could be a symmetry algebra of
PDE of the form (\ref{2.1}).

Thus there no are realizations of six-dimensional simple Lie
algebras by operators (\ref{n4}), which are symmetry algebras of
(\ref{2.1}).

The same assertion holds true for the simple Lie algebras of dimension
eight\ ($sl(3,{\bf R})$,\ $su(3)$,\ $su(2,1)$), which is
the next admissible dimension for real simple Lie algebras.

As\ $su^{*}(4) \sim so(5,1)$ and since the algebra\
$so(5,1)$\ contains\ $so(4)$, we conclude that the algebras\
$A_{n-1}\ (n>1)$\ have no realizations by operators of the form (\ref{n4}),
which generate symmetry algebras of (\ref{2.1}), except for those
given in Theorems 3.1 and  3.2.

There are also no realizations of the desired form for simple Lie
algebras of the type\ $D_n\ (n>1)$, since the lowest dimensional
algebras of this type\ ($so(4),\ so(2,2)$,\ $so^{*}(4)$) have no
realizations within the class (\ref{n4}) which could be symmetry
algebras of (\ref{2.1}).

By the same reasoning, we conclude that the realizations (\ref{n14}),
(\ref{n20})--(\ref{n24}) exhaust the set of all possible
realizations of the simple Lie algebras\ $B_n \ (n>1)$\ and\ $C_n\
(n\geq 1)$. Indeed, taking the least possible value of $n$ and
putting\ $n=2$\ we see that the algebras of the type\ $B_n$\
contain subalgebras that are isomorphic to\ $so(4),\
so(1,3)$. The same assertion for the simple Lie algebras of
the type $C_n\ (n\geq 1)$ follows from the relations:
\[
sp(2,{\bf R}) \sim so(3,2),\quad sp(1,1) \sim so(4,1),\quad sp(2)
\sim so(5),
\]
if we take into account that the algebras\ $so(3,2)$,\ $so(4,1)$,\
contain\ $so(3,1)$, and that the algebra\ $so(5)$\
contains\ $so(4)$.

To complete the proof we have to consider the exceptional simple
Lie algebras\ $G_2$,\ $F_4$,\ $E_6$,\ $E_7$,\ $E_8$.

We consider in detail the first two algebras, the remaining
algebras being treated in the same way.

A Lie algebra of the type\  $G_2$\ contains a compact real form\
$g_2$ and a non-compact real form\ $g_2'$. We also have\
$g_2\cap g_2' \sim su(2)\oplus su(2) \sim so(4)$, from which we
conclude that\ $G_2$\ has no realizations by operators of the form (\ref{n4})
which are symmetry operators of (\ref{2.1}).

A Lie algebra of the type\ $F_4$\ contains a compact real form\
$f_4$\ and two non-compact real forms\ $f_4',\ f_4''$. We also have\
$f_4'\cap f_4 \sim sp(3) \oplus su(2),\ f_4''\cap f_4
\sim so(9)$. Hence, it follows that the algebra\ $F_4$\ has no
realizations within the class of operators of the form (\ref{n4}) which are
admitted by PDE of the form (\ref{2.1}).

The theorem is proved.

\subsection{Equations invariant under semi-direct sums
of simple and solvable Lie algebras}

In order to describe equations of the form (\ref{2.1}) which are invariant
with respect to the Lie algebras that are semi-direct sums of
simple and solvable Lie algebras, we could follow the same
strategy as in the previous section. However, with Theorems
3.1--3.3 in hand, the most effective way is a direct application of the
Lie infinitesimal method in order to specify the arbitrary
functions of one variable, given in Theorems 3.1, 3.2, with the
aim of obtaining all the possible extensions of the algebras\
$so(3)$,\ $sl(2,{\bf R})$\ admitted by PDEs (\ref{2.1}). In this
way we will get all the possible equations of the form (\ref{2.1})
admitting Lie algebras which are semi-direct sums of simple and
solvable Lie algebras.

So we insert the corresponding forms of the functions\ $F$,\ $G$\
into invariance conditions (\ref{n5}) and then investigate
the consistency of the  system of determining equations which are obtained
in this way.
Substituting formulas (\ref{n15}) into the first equation of
(\ref{n5}) yields the following system of PDEs:
\begin{eqnarray*}
&(a)& 2b_x-\dot a-2c \ \tan u=0,\\ &(b)& b_u+c_x\sec ^2u=0,\\
&(c)& 2c_u-\dot a=0.
\end {eqnarray*}
It follows from $(c)$ that\ $c=\displaystyle{\frac{1}{2}}\dot
au+\tilde c(t,x)$. Then the compatibility requirement
for equations $(a)$ and $(b)$ gives\ $\dot a=0,\ \tilde
c_{xx} +\tilde c=0$,\ whence
\[
\dot a=0,\quad b=[f(t)\sin x-g(t)\cos x]\tan u+h(t), \quad
c=f(t)\cos x+g(t)\sin x,
\]
where\ $f,\  g,\  h$\ are arbitrary smooth functions\ of\ $t$.

Next, substituting the  expressions obtained for\ $F$\ and\ $G$\
into the second equation from (\ref{n15}) we see that\ $a \dot
{\tilde G}=0,\ \dot f=\dot g=\dot h=0$. Hence it follows that
extension of the symmetry algebra is only possible if\ $\tilde
G=\lambda,\ \lambda =const$.\ In this case, the maximal symmetry
algebra of the corresponding PDE is the four-dimensional Lie
algebra\ $so(3)\oplus L_1$,\ where\ $so(3)$\ is given in
(\ref{n14}), and\ $L_1=\langle \partial_t \rangle $.

We get similar results for PDEs invariant under the realizations
(\ref{n21}), (\ref{n23}) and (\ref{n24}). Namely, extension of the
symmetry algebra is only possible when\ $\tilde G=\lambda, \
\lambda ={\rm const}$.\ Moreover, the maximal invariance
algebras are the four-dimensional Lie algebras of the form\
$sl(2,{\bf R})\oplus L_1$,\ where\ $L_1= \langle \partial_u
\rangle $ when\ $sl(2,{\bf R})$\  given by (\ref{n21}),
and\ $L_1= \langle \partial_t \rangle $\ when \ $sl(2,{\bf
R})$\ is given by (\ref{n23}) or (\ref{n24}).

We turn now to the remaining case of PDE (\ref{2.1}) invariant with
respect to realization (\ref{n20}) of the algebra\  $sl(2,{\bf
R})$.\ Inserting the corresponding expressions for\ $F$\ and\ $G$\
into (\ref{n5}), we find the following equations:
\begin{eqnarray}
&&(A-2B\omega )\tilde F= (C+D\omega +B\omega ^2)\dot {\tilde F},
\label{n27} \\ &&(E+B\omega )\tilde G-(C+D \omega +B\omega ^2)
\dot {\tilde G}= \nonumber \\ &&\quad =K+L\omega +(M+N\omega
+P\omega ^2 + S\omega ^3) \tilde F - 2u(C+D\omega +B\omega ^2)\dot
{\tilde F}, \label{n28}
\end{eqnarray}
where
\begin{eqnarray}
A&=&2xb_x-x\dot a+4ub_u, \quad B=b_u, \nonumber \\
C&=&2xc-x^2c_x-2u(b+xc_u-xb_x)+4u^2b_u, \nonumber \\
D&=&b+xc_u-xb_x-4ub_u, \nonumber \\
K&=&-x^3c_t+2x^2ub_t+x^2uc_x+xu^2(c_u-2b_x+\dot a)-2u^3b_u,
\nonumber \\ L&=&-x^2b_t-xc+ub+ux(b_x-\dot a)+u^2b_u, \nonumber \\
E&=&2b+x(c_u-\dot a)-2ub_u, \nonumber \\
M&=&2uE-2xc+x^3c_{xx}-2x^2u(b_{xx}-2c_{xu})-4xu^2(2b_{xu}
-c_{uu})-8u^3b_{uu}, \nonumber \\
N&=&2ub_u+x^2(b_{xx}-2c_{xu})+4xu(2b_{xu}-c_{uu})+12u^2b_{uu},
\nonumber \\ P&=&-x(2b_{xu}-c_{uu})-6ub_{uu}, \quad S=b_{uu},
\nonumber \\ \dot {\tilde F}& =&\displaystyle{\frac{d\tilde
F}{d\omega}}, \quad \dot {\tilde G}=\displaystyle{\frac{d\tilde
G}{d\omega}}, \quad \omega =2u-xu_x. \nonumber
\end{eqnarray}

If\ $\tilde F$\ is an arbitrary function of\ $\omega $, then
we have\ $A=B=C=D=0$\. It follows that \ $b=\displaystyle{\frac{1}{2}}
x\dot a$,\ $c=x^2\tilde c(t)$. Equation (\ref{n28})
now takes the form
\[
K + L\omega =0,
\]
where
\[
K = -x^5\dot {\tilde c} + x^3u(\ddot a + 2\tilde c),\quad L =
-\displaystyle{\frac{1}{2}}x^3(\ddot a + 2\tilde c),
\]
so that,\ $\dot {\tilde c}=0$,\ $\ddot a + 2\tilde c = 0$. Thus
realization (\ref{n20}) of the algebra\ $sl(2,{\bf R})$\ is the
maximal invariance algebra of the corresponding PDE (\ref{2.1}).

Thus, extension of realization (\ref{n20}) is only possible
when not all of the coefficients\  $A, B, C, D$\ in (\ref{n27}) vanish as a
result
of (\ref{n27}), (\ref{n28}). In order to classify
all these cases we note that (\ref{n27}) is equivalent to the following
relation:
\begin{eqnarray}
&&(k-2m\omega )\tilde F=(n+p\omega +m\omega ^2) \dot {\tilde F},
\label{n29}
\end{eqnarray}
where the coefficients\ $k, m, n, p$ are constant. Indeed, since\
$\tilde F$\ is a function of\ $\omega$\ only, relation (\ref{n27})
can be valid if and only if all its coefficients have the form\ ${\rm
const.}\times R(t,x,u)$ with some non-vanishing function\ $R$. If
all the coefficients in (\ref{n27}) are equal to zero, then we get
the case of an arbitrary function\ $\tilde F$.\ Consequently,
extension of the invariance algebra is only possible when the
function\ $\tilde F(\omega )$\ satisfies an equation of the form
(\ref{n29}), where\ $k, m, n, p$,\ are  constants not vanishing
simultaneously.

Summing up, we conclude that the problem of the group classification of
PDEs (\ref{2.1}) invariant under realization (\ref{n20}) of the
algebra\ $sl(2,{\bf R})$,\ reduces to classifying all admissible
forms of the function\ $\tilde F$. Solving this problem requires
simple but very tedious computations and so we give only
the result, omitting the intermediate calculations. The admissible forms of
the functions\
$\tilde F(\omega)$\ are:
\begin{eqnarray}
\tilde F&=&1; \nonumber \\ \tilde F&=&\lambda \omega ^{\alpha };
\nonumber \\ \tilde F&=&\lambda \exp\omega ; \nonumber \\ \tilde
F&=&\lambda (\omega ^2+\alpha )^{-1}; \label{n33} \\ \tilde
F&=&\displaystyle{\frac{\lambda }{(\omega +\alpha )^2} \exp\left(
-\frac{2\alpha }{\omega +\alpha }\right)}; \nonumber \\ \tilde
F&=&\displaystyle{\frac{\lambda }{(\omega +\alpha )^2+ \beta ^2}
\exp\left(\frac{2\alpha }{\beta}\arctan \frac{\omega +\alpha
}{\beta}\right )}; \nonumber \\ \tilde
F&=&\displaystyle{\frac{\lambda }{(\omega +\alpha )^2-\beta ^2}
\exp\left| \frac{\omega +\alpha -\beta } {\omega +\alpha +\beta
}\right|^{\frac{\alpha}{\beta}}}, \nonumber
\end{eqnarray}
where\ $\{\alpha, \lambda, \beta\} \subset {\bf R},\ \alpha
\lambda \beta \not =0$.

On analyzing the above cases, we conclude that the only forms of
the function\ $\tilde F$ from the list (\ref{n33}) that provide
an extension of invariance algebra of the equation under study are:
\[
\tilde F=1,\quad \tilde F=\omega.
\]
We finally find that there exist five nonlinear equations of
the form (\ref{2.1}) invariant under four-dimensional algebras and
two nonlinear PDEs admitting five-dimensional Lie algebras. Below
we the give these equations together with their maximal symmetry
algebras\ $L_{\rm max}$.
\[
u_t=\displaystyle{\frac{\sec ^2u}{1+u_x^2\sec ^2u}}u_{xx}+
\displaystyle{\frac{1+2u_x^2\sec ^2u}{1+u_x^2\sec ^2u}}\tan u
+\lambda \sqrt{1+u_x^2\sec ^2u},\ \lambda \in {\bf R},
\]
$L_{\rm max}=so(3)\oplus L_1$,\ $so(3)$\ has the form (\ref{n14}),
$L_1=\langle
\partial_t\rangle ;$

\[
u_t=x^{-3}u_x^{-3}u_{xx}-\displaystyle{\frac{1}{4}}x^{-1}u_x
+3x^{-4}u_x^{-2}+\lambda x^{-3} u_x^{-1}, \ \lambda \in {\bf R},
\]
$L_{\rm max}=sl(2,{\bf R})\oplus L_1,$ \ $sl(2,{\bf R})$\ has the
form (\ref{n21}), $L_1=\langle
\partial_u\rangle ;$

\[
u_t=\displaystyle{\frac{u^2}{u^6+4u_x^2}}u_{xx}-\displaystyle
{\frac{10uu_x^2+u^7}{8u_x^2+2u^6}} + \lambda
u^{-2}\sqrt{u^6+4u_x^2}, \ \lambda \in {\bf R},
\]
$L_{\rm max}=sl(2,{\bf R})\oplus L_1,$ \ $sl(2,{\bf R})$\ has the
form (\ref{n23}), $L_1=\langle
\partial_t\rangle ;$

\[
u_t=\displaystyle{\frac{u^2}{u^6-4u_x^2}}u_{xx}+\displaystyle
{\frac{10uu_x^2-u^7}{8u_x^2-2u^6}} + \lambda u^{-2}\sqrt
{|u^6-4u_x^2|}, \ \lambda \in {\bf R},
\]
$L_{\rm max}=sl(2,{\bf R})\oplus L_1,$\ $sl(2,{\bf R})$\ has the
form (\ref{n24}), $L_1=\langle
\partial_t\rangle ;$

\[
u_t=\lambda[2u-xu_x]u_{xx}+[4\gamma -4\lambda
-1]x^{-2}u^2+[1+2\lambda -4\gamma] x^{-1}uu_x+\gamma u_x^2, \
\lambda \not =0, \ \gamma \in {\bf R},
\]
$L_{\rm max}=sl(2,{\bf R})\oplus L_1,$ \ $sl(2,{\bf R})$\ has the
form (\ref{n20}), $L_1=\langle x\,\partial_x+2u\,\partial_u\rangle
;$

\[
u_t=u^{-4}u_{xx}-2u^{-5}u_x^2,
\]
$L_{\rm max}=sl(2,{\bf R})\oplus L_{2.1},$ \ $sl(2,{\bf R})$\ has
the form (\ref{n22}), $L_{2.1}=\langle
4t\,\partial_t+u\,\partial_u,
\partial_t\rangle ;$

\[
u_t=u_{xx}+x^{-1}uu_x-x^{-2}u^2-2x^{-2}u,
\]
$L_{\rm max}=sl(2,{\bf R})\subset \hskip -3.8mm + L_{2.2},$ \
$sl(2,{\bf R})$\ has the form (\ref{n20}),\ $L_{2.2}=\langle
t\,\partial_x + [tx^{-1}(u+2)-x]\,
\partial_u, \partial_x+x^{-1}(u+2)\,\partial_u \rangle .$

The above formulas provide the full solution of the problem of
describing all PDEs of the type (\ref{2.1}) admitting symmetry Lie algebras
which are
semi-direct sums of semi-simple and solvable Lie algebras.

\subsection{Classification of equations (\ref{2.1}) invariant with
respect to solvable Lie algebras}

To complete the classification of invariant PDEs of the form
(\ref{2.1}) we have to construct all possible inequivalent
realizations of solvable Lie algebras within the class of
operators (\ref{n4}) which are invariance algebras of (\ref{2.1}).
First, we shall perform a preliminary classification: we
shall describe inequivalent PDEs (\ref{2.1}) admitting one-, two-
and three-dimensional solvable invariance algebras and then
proceed to classifying equations invariant with respect to higher
dimensional solvable Lie algebras.

\subsubsection{Preliminary classification}

Equations (\ref{2.1}) invariant with respect to one-dimensional
algebras have already been constructed, so that we can start by
considering two-dimensional solvable Lie algebras. As mentioned in
Section II.2 there are two inequivalent solvable Lie algebras
\begin{eqnarray}
A_{2.1}&:& [e_1, e_2] =0; \label{h5} \\ A_{2.2}&:& [e_1, e_2]
=e_2. \nonumber
\end{eqnarray}

As each of the above algebras contains the algebra\ $A_1$,\ when
studying realizations of two-dimensional Lie algebras we can take
as one of the basis operators either\ ${\partial\over \partial
t}$\ or\ ${\partial\over \partial x}$.\ Consider in more detail
the case of the algebra\ $A_{2.1}$.

Let\ $e_1 = \partial_t$\ and\ $e_2$\ be an operator of the form
(\ref{n4}). Then it follows from (\ref{h5}) that within a choice
of a basis of the algebra\ $A_{2.1}$\ we can put
\be \label{h6}
e_2 = b(x,u)
\partial_x +c(x,u) \partial_u.
\ee

Since operator (\ref{h6}) can be treated as the non-zero vector
field acting on smooth functions of\ $x, u$, we can choose\ $e_2 =
\partial_u$\ thus getting the realization\ $\langle \partial_t,
\partial_u \rangle $.

Now take the case\ $e_1 = \partial_x$ and\ $e_2$ is an operator
of the form (\ref{n4}). Using the commutation
relation (\ref{h5}) yields
\be \label{h7}
e_2 = a(t)\partial_t + b(t,u)
\partial_x +c(t,u) \partial_u.
\ee If\ $a \not =0$,\ we make the change of variables
\be
\label{h8} \overline{t} = T(t), \ \overline{x} = x +X(t,u), \ v =
U(t,u), \ \dot T \not =0, \ U_u \not =0, \ee
where\ $\dot T =a^{-1}$,\ $aX_t +x X_u +b =0;\ aU_t+cU_u=0,\ U_u
\not =0.$\ This reduces operator (\ref{h7}) to the form $e_2 =
\partial_{\overline{t}}$.

If in (\ref{h7})\ $a=0, \ c\not =0,$\ then we put\ $T = t$\ in
(\ref{h8}) and take as\ $X$ and $U$ solutions of the equations
$$
cX_u +b =0, \ \ c U_u =1,
$$
which reduces operator (\ref{h7}) to \ $e_2 = \partial_v$.

Finally, turning to the remaining case, when\ $a=c=0$\ in
(\ref{h7}).  There are transformations of the form
(\ref{h8}) which transform operator (\ref{h7})  to the form\
$e_2 = \overline{t} \partial_{\overline{x}}$\ (if\ $b_u= 0$)\
or to\ $e_2 = v \partial_{\overline{x}}$\ (if\ $b_u \not =0$).

Summing up we conclude that, up to equivalence defined by
transformations of the group\ $\cal E$,\ there are
four inequivalent realizations of the algebra\ $A_{2.1}:\ \langle
\partial_t, \partial_u \rangle$,\ $\langle \partial_x, \partial_u
\rangle$,\ $\langle \partial_x, t \partial_x \rangle$,\ $\langle
\partial_x, u \partial_x \rangle.$

The conditions (\ref{n5}) imply that the third realization, \ $\langle
\partial_x, t \partial_x \rangle$,\
cannot be an invariance algebra of PDEs of the form (\ref{2.1}).
The equation invariant under the fourth realization is
\be
\label{h9} u_t = u^{-2}_{x} F(t,x) u_{xx} +u_x G(t,u). \ee and is
linearizable by the change of variables $$ \overline{t} = t,
\ \ \overline{x} = u, \ \ v=x. $$

Thus we see that there exist two inequivalent realizations of the algebra\
$A_{2.1}$\ which are invariance algebras of nonlinear PDEs of the
form (\ref{2.1}):
\begin{eqnarray*}
A^1_{2.1} &=& \langle \partial_t, \partial_u \rangle; \\ A^2_{2.1}
&=& \langle \partial_x, \partial_u \rangle.
\end{eqnarray*}
The corresponding forms of the functions\ $F$, $G$\ are given in
Table 1.

The same reasoning gives all inequivalent realizations of
the abstract Lie algebra\ $A_{2.2}$. The full list of these
contains three realizations which are admitted by PDEs of the
form (\ref{2.1}):
\begin{eqnarray*}
A^1_{2.2} &=& \langle - t \partial_t - x\partial_x, \partial_t
\rangle;\\ A^2_{2.2} &=& \langle - t \partial_t - x\partial_x,
\partial_x \rangle;\\ A^3_{2.2} &=& \langle - x \partial_x -
u\partial_u, \partial_x \rangle.
\end{eqnarray*}
The corresponding forms of the functions\ $F$ and $G$ are given in
Table 1.

Let us note that provided the functions\ $\tilde F, \tilde G$\
are arbitrary, the corresponding realizations of the
two-dimensional Lie algebras are maximal invariance algebras of
these equations.
\newpage

{\bf  Table 1.}\ {\it Invariance of (\ref{2.1}) under
two-dimensional solvable Lie algebras} \vspace{3mm}

\begin{center}
\begin{tabular}{|c|c|c|} \hline
Algebra& $F$&$G$ \\ \hline & &  \\ $A^1_{2.1}$&${\tilde F}(x,
u_x)$&${\tilde G} (x,u_x)$ \\ [2mm ] & &  \\ $A^2_{2.1}$&${\tilde
F}(t, u_x)$&${\tilde G} (t,u_x)$ \\ [2mm]  & &  \\ $A^1_{2.2}$&$ x
{\tilde F}(u, \omega)$&$x^{-1}{\tilde G} (u, \omega), \ \omega = x
u_x$ \\ [2mm]  & &  \\ $A^2_{2.2}$&$ t {\tilde F}(u,
\omega)$&$t^{-1}{\tilde G} (u, \omega), \ \omega = t u_x$ \\ [2mm]
 & & \\ $A^3_{2.2}$&$ u^2{\tilde F}(t, u_x)$&$ u {\tilde G}
(t,u_x)$ \\ & & \\ \hline
\end{tabular}
\end{center}
\vspace{2mm}

We begin the search for realizations of three-dimensional solvable
Lie algebras\ $A_3 = \langle e_1$, $e_2, e_3 \rangle$\ by
considering decomposable algebras\ $A_{3.1}, A_{3.2}$.

Evidently, in order to get all the possible realizations of these
algebras within the class of operators (\ref{n4}), it suffices to
extend the  realizations already known for the two-dimensional
algebras\ $A^i_{2.1} = \langle e_1, e_2 \rangle \ (i=1,2)$ (for
the algebra\ $A_{3.1}$)\ and\ $A^i_{2.2} = \langle e_1, e_2
\rangle \ (i=1,2, 3)$ (for the algebra\ $A_{3.2}$).\ This follows
from the definition of decomposable solvable Lie algebras.
As a result we get one realization of the algebra\ $A_{3.1}$\ and six
inequivalent realizations of the algebra\ $A_{3.2}$,\ which are admissible as
invariance algebras of PDEs (\ref{2.1}):
\begin{eqnarray*}
A^1_{3.1} &=& \langle \partial_t, \partial_u, \partial_x \rangle
;\\ A^1_{3.2} &=& \langle -t \partial_t-x \partial_x, \partial_t,
\partial_u \rangle ;\\ A^2_{3.2} &=& \langle -t \partial_t-u
\partial_u, \partial_t, x u \partial_u \rangle ;\\ A^3_{3.2} &=&
\langle -t \partial_t-u \partial_u, \partial_u, t\partial_t+x
\partial_x \rangle ;\\ A^4_{3.2} &=& \langle -t \partial_t-x
\partial_x, \partial_x, \partial_u \rangle ;\\ A^5_{3.2} &=&
\langle -x \partial_x-u \partial_u, \partial_u,  \partial_t
\rangle ;\\ A^6_{3.2} &=& \langle -x \partial_x-u \partial_u,
\partial_u,  tx\partial_x \rangle .
\end{eqnarray*}

The explicit forms of the invariant equations (\ref{2.1}) are
determined by the forms of the functions\ $F$, $G$ which are given
in Table 2, where\ $\tilde F$, $\tilde G$\ are arbitrary smooth
functions.

One can verify by direct computation that the
realizations given above are the maximal invariance algebras of the
corresponding
equations, provided the functions\ $\tilde F$\ and\ $\tilde G$\ are
arbitrary smooth functions.

As mentioned in Section II.2, there are seven abstract
non-isomorphic non-decomposable Lie algebras. All of them contain
the two-dimensional commutative ideal\ $A_{2.1} = \langle e_1, e_2
\rangle$.\ Thus, to construct their realizations within the
class of operators under consideration, it suffices to describe all
the possible extensions of the realizations of\ $A_{2.1}$\ with
the operator\ $e_3$\ of the form (\ref{n4}). Moreover, we
have to consider both the realizations\ $A^i_{2.1} = \langle e_1,
e_2 \rangle \ (i=1,2),$\ and\ ${\tilde A}^i_{2.1} = \langle
{\tilde e}_1, {\tilde e}_2 \rangle \ (i=1,2),$\ where\ $\tilde e_1
= e_2, \tilde e_2 = e_1.$ We will consider in more detail the
procedure for constructing realizations of the Weyl algebra\
$A_{3.3}$,\ which is a nilpotent Lie algebra.
\newpage

\noindent {\bf  Table 2.}\ {\it Invariance of (\ref{2.1}) under
three-dimensional decomposable solvable Lie algebras}
\vspace{3mm}

\begin{center}
\begin{tabular}{|c|c|c|} \hline
Algebra& $F$&$G$ \\ \hline & & \\ $A^1_{3.1}$&${\tilde
F}(u_x)$&${\tilde G} (u_x)$  \\  & & \\ $A^1_{3.2}$&$x {\tilde
F}(\omega)$&$ x^{-1}{\tilde G} (\omega), \ \omega = x u_x$
\\ & & \\ $A^2_{3.2}$&$ u^{-1} e^{x \omega} {\tilde
F}(x)$&$e^{x \omega}[{\tilde G} (x)-\omega^2 {\tilde F}(x)],
 \ \omega = u^{-1} u_x$   \\
& & \\ $A^3_{3.2}$&$t^{-1} x^{2} {\tilde F}(\omega)$&$
x^{-1}{\tilde G} (\omega), \ \omega = t^{-1} x^2 u_x$   \\ & &
\\ $A^4_{3.2}$&$t{\tilde F}(\omega)$&$ t^{-1} {\tilde G} (\omega),
\ \omega = t u_x$   \\ & & \\ $A^5_{3.2}$&$x^2 {\tilde F}(u_x)$&${
x \tilde G} (u_x)$   \\ & & \\ $A^6_{3.2}$&$ x^{2} {\tilde
F}(t)$&$ x t^{-1}u_x \ln | u_x |+ x u_x {\tilde G}(t)$  \\ & & \\
\hline
\end{tabular}
\end{center}
\vspace{2mm}

We begin with the realization\ $A^1_{2.1}$.\ If\ $e_1 =
\partial_t, e_2 = \partial_u$,\ then it follows from the commutation
relation\ $[e_2, e_3] = e_1,$\ where\ $e_3$\ is of the form
(\ref{n4}), that the equation\ $b_u \partial_x+c_u \partial_u =
\partial_t$\ holds true. Since this equation cannot be satisfied
for any choice of the functions\ $a, b, c$\ contained in\ $e_3$,
this realization cannot be extended to that of a three-dimensional
solvable Lie algebra. Next, if\ $e_1 = \partial_u, e_2 =
\partial_t$,\ then\ $e_3 = {\tilde b}(x)\partial_x + [t+{\tilde
c}(x)]\partial_u$, and we obtain, up to equivalence under \ $\cal
E$,\ the following three realizations of the algebra\ $A_{3.3}$:
\begin{eqnarray*}
&& \langle \partial_u, \partial_t, \partial_x+t \partial_u
\rangle, \\ && \langle \partial_u, \partial_t, t \partial_u
\rangle, \\ && \langle \partial_u, \partial_t, (t+x) \partial_u
\rangle.
\end{eqnarray*}

Checking conditions (\ref{n5}) we find that the second realization
from the above list cannot be an invariance algebra of PDE
(\ref{2.1}). Moreover,  the equation admitting the third
realization is necessarily linear.

In studying the realization\ $A^2_{2.1}$\ we have to take into
account the existence of two possibilities. The first possibility
is\ $e_1 = \partial_x, e_2 = \partial_u$\ and the second one is\
$e_1 = \partial_u, e_2 = \partial_x$. However, the above
realizations are transformed one into another by the change of
variables
\be \label{h11} \overline{t} = t,\ \ \overline{x} = u, \
\ v =x. \ee
So we may consider without loss of generality the second
realization only. Performing the necessary computations yields a
realization of the algebra\ $A_{3.3}$\ which can be admitted by a
nonlinear equation of the form (\ref{2.1}):\ $\langle
\partial_u,\partial_x,\partial_t+x \partial_u \rangle.$

We conclude that there are two inequivalent
realizations of the algebra\ $A_{3.3}$,\ which are invariance
algebras of nonlinear PDEs from the class (\ref{2.1}),
\begin{eqnarray*}
A^1_{3.3} &=& \langle \partial_u, \partial_t,
t\partial_u+\partial_x \rangle ;\\ A^2_{3.3} &=& \langle
\partial_u, \partial_x, t\partial_x+x\partial_u \rangle .
\end{eqnarray*}
The forms of the functions\ $F$ and $G$\ defining the
corresponding nonlinear heat conductivity equations are given in
Table 3.

\newpage

{\bf Table 3.}\ {\it Invariance of (\ref{2.1}) with respect to the
Weyl algebra} \vspace{3mm}

\begin{center}
\begin{tabular}{|c|c|c|} \hline
Algebra& $F$&$G$ \\ \hline & & \\ $A^1_{3.3}$&${\tilde
F}(u_x)$&$x+{\tilde G} (u_x)$
\\ & & \\ $A^2_{3.3}$&${\tilde F}(t)$&$-\frac{1}{2}
u^2_x+{\tilde G}(t)$  \\ & & \\ \hline
\end{tabular}
\end{center}

The remaining non-decomposable solvable Lie algebras are
treated in an analogous way. We present below those of their
inequivalent realizations which are admitted by nonlinear PDEs of
the form (\ref{2.1}). In Table 4 we give the various forms of
the functions\ $F$, $G$\ defining the forms of the invariant
equations.
\begin{eqnarray*}
A^1_{3.4}&=& \langle \partial_u, \partial_t, t \partial_t+x
\partial_x+[t+u] \partial_u\rangle ;\\ A^2_{3.4}&=& \langle
\partial_u, \partial_t, t \partial_t+[t+u] \partial_u\rangle ;\\
A^3_{3.4}&=& \langle \partial_x, \partial_u, 2t \partial_t+(x+u)
\partial_x+u \partial_u\rangle ;\\ A^4_{3.4}&=& \langle
\partial_x, \partial_u, (x+u) \partial_x+u \partial_u\rangle ;\\
A^1_{3.5}&=& \langle \partial_t, \partial_u, t \partial_t+x
\partial_x+u \partial_u\rangle ;\\ A^2_{3.5}&=& \langle
\partial_t, \partial_u, t \partial_t+u \partial_u\rangle ;\\
A^3_{3.5}&=& \langle \partial_x, \partial_u, 2t \partial_t+x
\partial_x+u \partial_u\rangle ;\\ A^1_{3.6}&=& \langle
\partial_t, \partial_u, t \partial_t+x \partial_x-u
\partial_u\rangle ;\\ A^2_{3.6}&=& \langle \partial_t, \partial_u,
t \partial_t-u \partial_u\rangle ;\\ A^3_{3.6}&=& \langle
\partial_x, \partial_u, t \partial_t+x \partial_x-u
\partial_u\rangle ;\\ A^4_{3.6}&=& \langle \partial_x, \partial_u,
x \partial_x-u \partial_u\rangle ;\\ A^1_{3.7}&=& \langle
\partial_u, \partial_t, q t \partial_t+x \partial_x+u
\partial_u\rangle \ (q\not=0, \pm 1) ;\\ A^2_{3.7}&=& \langle
\partial_u, \partial_t, q t \partial_t+u \partial_u\rangle \
(q\not=0, \pm 1) ;\\ A^3_{3.7}&=& \langle \partial_x, \partial_u,
t \partial_t+x \partial_x+qu \partial_u\rangle \ (0<|q|<1) ;\\
A^4_{3.7}&=& \langle \partial_x, \partial_u, x \partial_x+qu
\partial_u\rangle \ (0<|q|<1) ;\\ A^1_{3.8}&=& \langle \partial_x,
\partial_u, \partial_t+u \partial_x-x \partial_u\rangle ;\\
A^2_{3.8}&=& \langle \partial_x, \partial_u, u \partial_x-x
\partial_u\rangle ;\\ A^1_{3.9}&=& \langle \partial_x, \partial_u,
\partial_t+(u+qx) \partial_x+(qu-x) \partial_u\rangle \ (q>0);\\
A^2_{3.9}&=& \langle \partial_x, \partial_u,  (u+qx)
\partial_x+(qu-x) \partial_u\rangle \ (q>0).
\end{eqnarray*}

Let us note that if the functions\ $\tilde F$, $\tilde G$\
from Tables 3, 4 are arbitrary, then the corresponding realizations are
the maximal invariance algebras of the equations obtained.

\subsubsection{Complete classification of nonlinear PDEs
(\ref{2.1}) invariant with respect to solvable Lie algebras}

The next step of our approach to the group classification of nonlinear
PDEs of the form (\ref{2.1}) is to describe equations which are invariant
under four-dimensional solvable Lie algebras.

\newpage
{\bf  Table 4.}\ {\it Invariance of (\ref{2.1}) under
non-decomposable three-dimensional solvable Lie algebras}
\vspace{3mm}

\begin{center}
\begin{tabular}{|c|c|c|} \hline
Algebra& $F$&$G$ \\ \hline & & \\ $A^1_{3.4}$&$x {\tilde
F}(u_x)$&${\tilde G} (u_x)+\ln | x |$ \\ & & \\
$A^2_{3.4}$&$u^{-1}_x {\tilde F}(x)$&${\tilde G} (x)+\ln | u_x |$
\\ & & \\ $A^3_{3.4}$&$u^{-2}_x {\tilde F}(\omega)$&$u_x
e^{-\frac{1}{u_x}}{\tilde G} (\omega), \ \omega = 2 u^{-1}_x-\ln |
t |$  \\ & & \\ $A^4_{3.4}$&$u^{-2}_x {\tilde F}(t) \exp(2
u^{-1}_x)$&$u_x {\tilde G}(t)\exp(u^{-1}_x)$  \\ & & \\
$A^1_{3.5}$&$ x {\tilde F}(u_x) $&${\tilde G} (u_x)$  \\ & &
\\ $A^2_{3.5}$&$ u^{-1}_x {\tilde F}(x)$&${\tilde G} (x)$   \\
& & \\ $A^3_{3.5}$&$ {\tilde F} (u_x) $&$|t|^{-\frac{1}{2}}{\tilde
G} (u_x)$  \\ & & \\ $A^1_{3.6}$&$x {\tilde F}(\omega)$&$ x^{-2}
{\tilde G} (\omega), \ \omega = x^2 u_x$  \\ & & \\ $A^2_{3.6}$&$
u_x {\tilde F}(x)$&$u^2_x {\tilde G} (x)$  \\ & & \\
$A^3_{3.6}$&$t {\tilde F}(\omega)$&$ t^{-2} {\tilde G} (\omega), \
\omega = t^2 u_x$  \\ & & \\ $A^4_{3.6}$&$ u^{-1}_x {\tilde
F}(t)$&$\sqrt{|u_x|} {\tilde G} (t)$  \\ & & \\ $A^1_{3.7}$&$
|x|^{2-q}{\tilde F} (u_x) $&$|x|^{1-q}{\tilde G} (u_x) \ (q=\not
=0, \pm 1)$  \\ & & \\ $A^2_{3.7}$&$ |u_x|^{-q}{\tilde F} (x) $&$
|u_x|^{1-q}{\tilde G}(x),q \not =0, \pm 1$  \\ & &
\\ $A^3_{3.7}$&$ t{\tilde F} (\omega) $&$|t|^{q-1}{\tilde G}
(\omega), \ \omega=|t|^{1-q} u_x ( 0<|q|<1)$ \\ & & \\
$A^4_{3.7}$&$ |u_x|^{\frac{2}{q-1}} {\tilde
F}(t)$&$|u_x|^{\frac{q}{q-1}}{\tilde G} (t) \ ( 0<|q|<1)$
\\ & & \\
$A^1_{3.8}$&$(1+u^2_x)^{-1}{\tilde F}(\omega)$&$\sqrt{1+u^2_x}
{\tilde G}(\omega) , \ \omega = t +\arctan u_x$  \\ & & \\
$A^2_{3.8}$&$(1+u^2_x)^{-1}{\tilde F}(t)$&$\sqrt{1+u^2_x} {\tilde
G}(t) $  \\ & & \\ $A^1_{3.9}$&$\frac{\textstyle{ \exp (-2 q
\arctan u_x){\tilde
F}(\omega)}}{\textstyle{1+u^2_x}}$&$\sqrt{1+u^2_x}\exp(-q \arctan
u_x) {\tilde G}(\omega) , $  \\ & & $\ \omega = t +\arctan u_x \
(q>0)$  \\ & & \\ $A^2_{3.9}$&$\frac{\textstyle{\exp(-2 q \arctan
u_x){\tilde F}(t)}}{\textstyle{1+u^2_x}}$&$\sqrt{1+u^2_x}\exp(- q
\arctan u_x) {\tilde G}(t) , \ (q>0) $  \\ & & \\ \hline
\end{tabular}
\end{center}

As we mentioned in Section II.2, there are ten decomposable and ten
non-decomposable, non-isomorphic, solvable four-dimensional Lie
algebras. Since nonlinear equations of the form (\ref{2.1}) which
admit three-dimensional solvable algebras contain arbitrary
functions of one argument, it is only natural to expect that
PDEs admitting four-dimensional algebras will depend on arbitrary
parameters at most. In other words, the arbitrary functions in
question will take specific forms dictated by the extension of symmetry group.
This is, indeed, the case for all the invariant PDEs except for the
equation \be \label{a45} u_t =
F(u_x) u_{xx}. \ee Group classification of PDEs of the form
(\ref{a45}) has been carried out in \cite{ah:} and we give below
the results obtained in the form of theorem.

\begin{thm} \mbox{\bf (\cite{ah:})}\ Provided\ $F$\ is an
arbitrary smooth function, the maximal invariance algebra admitted
by (\ref{a45}) is the four-dimensional Lie algebra
$$A^1_{3.1} + \hskip -3.8mm \supset \langle 2t \partial_t +x
\partial_x +u \partial_u \rangle.$$

An extension of the symmetry algebra of PDE (\ref{a45}) is only
possible for the three cases given below:
\begin{eqnarray*}
F = \exp u_x &:& e_5=t \partial_t -x \partial_u; \\ F = u^n_x &:&
e_5=nt \partial_t - u \partial_u, \ \ n \ge -1, \ \ n\not =0;\\ F
= \frac{\exp(n \arctan u_x)}{1+u^2_x} &:& e_5=nt \partial_t - u
\partial_x- x \partial_u , \ \ n \ge 0.
\end{eqnarray*}
\end{thm}

In view of this result, we will exclude from further consideration
equations which are equivalent to an equation of the form
(\ref{a45}).

Consider first the decomposable solvable four-dimensional Lie
algebras
\begin{eqnarray*}
&& 4A_1 = A_{3.1} \oplus A_1, \hskip 15mm A_{3.2} \oplus A_1, \\
&& 2 A_{2.2} = A_{2.2} \oplus A_{2.2}, \hskip 10mm A_{3.i} \oplus
A_1 \ \ \ (i=3,4, \ldots, 9).
\end{eqnarray*}

On analyzing extensions of the realization\ $A^1_{3.1}$\ for the
algebra\ $4A_1$\ and of the realizations\ $A^i_{3.2} \ (i=1,
\ldots, 6)$\ by an operator\ $e_4$\ of the form (\ref{n4}), we
conclude that there are no realizations of the algebras\ $4A_1$\
and\ $A_{3.2} \oplus A_1$\ which could be invariance algebras of
PDEs of the form (\ref{2.1}).

Studying realizations of the algebra\ $2A_{2.2}$\ yields four
inequivalent realizations admitted by PDEs from the class
(\ref{2.1}). We give these realizations below, as well as the
corresponding forms of the functions\ $F$,\ $G$.
\begin{eqnarray*}
2 A^1_{2.2} &= &A^1_{3.2} + \hskip -3.8mm \supset \langle -u
\partial_u +kx \partial_x \rangle
 \  \  (k \not =0): \\
&& F = \lambda x | \om |^{-k}, \ \ \ G = \beta x^{-1}
|\omega|^{1-k}, \  \  \lambda \not =0, \beta \in R, \om = x u_x;\\
2 A^2_{2.2} &= &A^2_{3.2} + \hskip -3.8mm  \supset \langle x
\partial_x \rangle : \\ && F = \lambda x^2 u^{-1} \exp \om, \ \ \
G = (\beta -\lambda \om^2)\exp \om, \  \  \lambda \not =0, \beta
\in R, \om = x u^{-1} u_x;\\ 2 A^3_{2.2} &= &A^4_{3.2} + \hskip
-3.8mm  \supset \langle -u \partial_u +kt \partial_t \rangle
 \  \  (k \not =0,1): \\
&& F = \lambda t | \om |^{\frac{2k}{1-k}}, \ \ \ G = \beta t^{-1}
|\omega|^{\frac{1}{1-k}},
 \ \ \om = t u_x, \ \ \lambda \not =0,
\beta \in R;\\ 2 A^4_{2.2} &= &A^4_{3.2} + \hskip -3.8mm  \supset
\langle -u \partial_u +t \partial_x \rangle
 : \\
&& F = \lambda t, \ \ \ G = u_x \ln |t u_x| + \beta u_x,
 \ \ \ \lambda \not =0, \ \ \beta \in R.
\end{eqnarray*}

Now, in order to complete group classification of the  PDEs given above,
one has to compute their maximal invariance algebras. To this end,
it is necessary to solve the determining equations (\ref{n5}) for
each choice of the functions\ $F$, $G$. Note, that we have
simplified the forms of the functions\ $F, G$\ with the use of
transformations from the corresponding equivalence groups which
are subgroups of\ ${\cal E}$. As a result, we get the following
simplified forms of the above invariant equations:
\begin{eqnarray}
&& 2 A^1_{2.2}: u_t= |x|^{1-k} |u_x|^{-k} u_{xx}+ \beta |x|^{-k}
|u_x|^{1-k}, \ \ \beta \in R, \ \ k\not =0; \label{a47}  \\ && 2
A^2_{2.2}: u_t= x^{2}u^{-1} \exp (\om) u_{xx}+ (\beta-\om^2) \exp
\om, \ \ \omega = x u^{-1} u_x, \ \ \beta \in R;  \label{a48}  \\
&& 2 A^3_{2.2} : u_t = \pm
|t|^{\frac{k+1}{1-k}}|u_x|^{\frac{2k}{1-k}} u_{xx} + \epsilon
|t|^{\frac{k}{1-k}} |u_x|^{\frac{1}{1-k}},  \ \ \epsilon=0,1, \ \
k \not =0,1;  \label{a49} \\ && 2 A^4_{2.2} : u_t =\lambda t
u_{xx} +u_x \ln|t u_x|, \ \ \ \lambda \not =0. \label{a50}
\end{eqnarray}

Inserting the functions\ $F$, $G$\ defining the above PDEs
(\ref{a47})--(\ref{a50}) into the determining equations (\ref{n5}) and
analyzing the equations obtained, we arrive at the following
conclusions.
\begin{enumerate}
\item If\ $k \not =0,2, \ \beta \not = \frac{k-1}{k-2}$\
or\ $k=2, \beta \not =\frac{5}{4}$\ in (\ref{a47}), the realization\
$2A^1_{2.2}$\ is the maximal invariance algebra of the nonlinear
heat conductivity equation (\ref{a47}). If\ $k=2, \beta
=\frac{5}{4}$,\ then the maximal invariance algebra of the
equation in question is five-dimensional. Its basis if formed by
the operators of the realization\ $2A^1_{2.2}\ \ (k=2)$\ and the
operator\ $4 xu \partial_x - u^2 \partial_u$. However, this
algebra is isomorphic to the Lie algebra\ $sl(2,R) \oplus
A_{2.2}$\ and the change of variables $$ \overline{t} = t, \hskip
10mm \overline{x} = u, \hskip 10mm v = \alpha |x|^{\frac{1}{4}}, \
\ \alpha \not =0 $$ transforms its basis operators to become basis
operators of the realization of\ $sl(2,R) \oplus L_{2.1}$,\ where\
$sl(2,R)$\ is the realization (\ref{n22}) and\  $L_{2.1} = \langle
4t \partial_t +u \partial_u, \partial_t \rangle.$ Thus equation
(\ref{a47}) with $k=2, \ \beta=\frac{5}{4}$ is equivalent to
an invariant PDE obtained in the previous section.

Finally, if\ $k\not =0,2$\ and\  $\beta = \frac{k-1}{k-2}$\ in
equation (\ref{a47}), then the latter is transformed by
transformations from the group\ $\cal E$\ to the form (\ref{a45}).
\item If\ $\beta \not =-2$\ in equation (\ref{a47}), the
realization\ $2 A^2_{2.2}$\ is the maximal invariance algebra of
this equation. Given the condition\ $\beta =-2$, the maximal
invariance algebra is the five-dimensional Lie algebra spanned by
the operators
$$ \langle \partial_t, -xu \partial_u, x^2 \partial_x +\ln |x^2 u|
xu \partial_u, 2t \partial_t+2u \partial_u-x \partial_x, t
\partial_t +u \partial_u \rangle.$$
The change of variables $$ \overline{t} = t, \hskip 10mm
\overline{x} = -x^{-1}, \hskip 10mm v = x^{-1}\ln |u| +2 x^{-1}
(1+\ln|x|) $$ reduce equation (\ref{a48}) with\ $\beta =-2$\ to
the equation $$ v_{\overline{t}} = \exp(v_{\overline{x}})
v_{\overline{x}\overline{x}},$$
which is contained in the class of PDEs (\ref{a45}).
\item The realization\ $2A^3_{2.2} \ (k \not =0,1)$\ is the
maximal invariance algebra of PDE (\ref{a49}), provided\ $\epsilon
=1$. If\ $\epsilon =0$,\ then its maximal invariance algebra is
the five-dimensional Lie algebra spanned by the operators
$$2A^3_{2.2} \ \ (k \not =0,1) \subset \hskip -3.8mm + \langle
|t|^{\frac{1+k}{k-1}} \partial_t \rangle.$$
However, with this choice of\ $\varepsilon$,\ equation (\ref{a49})
is reduced through the change of variables
$$\overline{t} = \frac{1}{2}(1-k) |t|^{\frac{2}{1-k}}, \hskip 10mm
\overline{x} = x, \hskip 10mm v =u,$$
to equation
$$v_{\overline{t}} = \pm |v_{\overline{x}}|^{\frac{2k}{1-k}}
v_{\overline {x}\overline{x}},$$
which belongs to the class of PDEs (\ref{a45}).
\item The realization\ $2A^4_{2.2}$\ is the maximal invariance
algebra of PDE (\ref{a50}). Analyzing the algebra\ $A_{3.3} \oplus
A_1$\ we find that it has no realizations which are
admissible for PDEs (\ref{2.1}). Next, we get a realization\
$A^2_{3.5} \oplus \langle \partial_x\rangle$\ of the algebra\
$A_{3.5}\oplus A_1$\ but the corresponding invariant equation
$$u_t = u^{-1}_{x} u_{xx}$$
belong to the class of PDEs (\ref{a45}). Studying realizations of
the algebra\ $A_{3.7} \oplus A_1$\ we get the nonlinear heat
conductivity equation
$$u_t = u^{-2}_{x} u_{xx}+u^{-1}_x,$$
whose maximal invariance algebra is the five-dimensional algebra
having the following basis elements:
$$A^1_5 = A^2_{3.7} \ (q=2) \oplus \langle
\partial_x, e^x \partial_x \rangle. $$
A similar analysis of the remaining decomposable solvable
four-dimensional algebras yields eight inequivalent realizations
that are maximal invariance algebras of nonlinear PDEs of the form
(\ref{2.1}). We list the nonlinear PDEs (\ref{2.1}) whose
maximal invariance algebras are four-dimensional decomposable
solvable Lie algebras in Table 5.
\end{enumerate}

We turn now to non-decomposable algebras. There are ten
non-isomorphic non-de\-com\-pos\-able solvable four-dimensional
Lie algebras\ $A_4 = \langle e_i| i=1,2,3,4 \rangle$\ (the full
list is given in Section II.2). Their structure implies that
studying realizations of these algebras can be carried out by
extension of the (already known) realizations of three-dimensional
solvable algebras\ $A_3 = \langle e_1, e_2, e_3 \rangle$\ by the
operator\ $e_4$\ of the form (\ref{n4}). Moreover, one must use
the following extension scheme: $A_{4.i} = A_{3.1} +\hskip -3.8mm
\supset \langle e_4 \rangle$ \ \ $(i=1, \ldots, 6),$ \ $A_{4.i} =
A_{3.3}+\hskip -3.8mm \supset \langle e_4 \rangle$  \ \
$(i=7,8,9)$, \ \ $A_{4.10} = A_{3.5} +\hskip -3.8mm \supset
\langle e_4 \rangle.$

There exists only one realization of the algebra\ $A_{3.1}$,\
and it is the maximal invariance algebra of the PDE \be \label{a51} u_t
=F(u_x) u_{xx} +G(u_x), \ee
so that nonlinear PDEs invariant under
realizations of the algebras\ $A_{4.i} \ (i=1, \ldots, 6)$\ must
belong to the class of equations (\ref{a51}).

Direct computation shows that the algebra\ $A_{4.1}$ has no
realizations that are admitted by PDEs (\ref{2.1}). For the
remaining abstract Lie algebras from the class under study we get
seven realizations which are invariance algebras of nonlinear PDEs
of the form (\ref{2.1}).
\begin{eqnarray*} && A^1_{4.2} =
A^1_{3.1}+\hskip -3.8mm \supset \langle qt
\partial_t +x \partial_x +(u+x) \partial_u \rangle \ \ (q \not
=0,1); \\ && A^2_{4.2} = A^1_{3.1}+\hskip -3.8mm \supset \langle t
\partial_t +(t+x) \partial_x +qu \partial_u \rangle \ \ (q \not
=0,1); \\ && A^1_{4.3} = A^1_{3.1}+\hskip -3.8mm \supset \langle t
\partial_t +x\partial_u \rangle; \\ && A^2_{4.3} =
A^1_{3.1}+\hskip -3.8mm \supset \langle t \partial_x +u\partial_u
\rangle; \\ && A^1_{4.4} = A^1_{3.1}+\hskip -3.8mm \supset \langle
t \partial_t +(t+x) \partial_x +(x+u)\partial_u \rangle; \\ &&
A^1_{4.5} = A^1_{3.1}+\hskip -3.8mm \supset \langle t \partial_t +
px \partial_x +qu\partial_u \rangle \ \ (p<q, \ p \cdot q \not =0;
\ p,q, \not =1); \\ && A^1_{4.6} = A^1_{3.1}+\hskip -3.8mm \supset
\langle qt \partial_t +(px +u)\partial_x +(pu-x)\partial_u \rangle
\ \ ( q \not =0; \ p\ge 0).
\end{eqnarray*}
The corresponding forms of the functions\ $F$, $G$ defining
invariant equations (\ref{2.1}) are given in Table 6. Note that,
for the sake of completeness, we give in Table 6 equation
(\ref{a45}), whose maximal invariance algebra for arbitrary\
$F$\ is\ $$ A^2_{4.5} = A^1_{3.1}+\hskip -3.8mm \supset
\langle t
\partial_t +\frac{1}{2}x \partial_x +\frac{1}{2}u\partial_u
\rangle. $$

\newpage
\begin{center}
{\bf Table 5.}\ {\it Invariance of (\ref{2.1}) under decomposable
four-dimensional solvable algebras} \vspace{3mm}

\begin{tabular}{|c|c|c|} \hline
Algebra& $F$&$G$ \\ \hline & & \\ $2A^1_{2.2}, \  \ (k \not =0,2)$
& $|x|^{1-k} |u_x|^{-k}$&$ \beta |x|^{-k} |u_x|^{1-k}, \beta \not
=\frac{k-1}{k-2}$ \\ & & \\ $2A^1_{2.2}\ (k=2)$ & $x^{-1}
u^{-2}_x$&$ \beta x^{-2} u^{-1}_x, \beta \not =\frac{5}{4}$ \\ & &
\\ $2A^2_{2.2}$ & $x^2 u^{-1}\exp \om$&$ (\beta-\om^2) \exp \om, \
\om =x u^{-1}u_x, \beta \not =-2$ \\ & & \\ $2A^3_{2.2}\ (k\not
=0,1)$ & $\pm |t|^{\frac{k+1}{1-k}} |u_x|^{\frac{2k}{1-k}}$&$
|t|^{\frac{k}{1-k}} |u_x|^{\frac{1}{1-k}}$ \\ & & \\ $2A^4_{2.2}$
& $\lambda t, \ \ \lambda \not =0$&$ u_x \ln|t u_x|$ \\ & & \\
$A^2_{3.4}\oplus \langle \partial_x \rangle $ &
$u^{-1}_x$&$\ln|u_x|$ \\ & & \\ $A^4_{3.4}\oplus \langle
\partial_t \rangle $ & $u^{-2}_x \exp (2 u^{-1}_x)$&$ u_x
\exp(u^{-1}_x)$ \\ & & \\ $A^2_{3.6}\oplus \langle \partial_x
\rangle $ & $u_x$&$u^2_x$ \\ & & \\ $A^4_{3.6}\oplus \langle
\partial_t \rangle $ & $u^{-1}_x$&$\sqrt{|u_x|}$ \\ & & \\
$A^2_{3.7}\oplus \langle \partial_x \rangle \ (q\not =0, \pm 1, 2)
$ & $|u_x|^{-q}$&$|u_x|^{1-q}$ \\ & & \\ $A^4_{3.7}\oplus \langle
\partial_t \rangle \ (0<|q|<1) $ &
$|u_x|^{\frac{2}{q-1}}$&$|u_x|^{\frac{q}{q-1}}$ \\ & & \\
$A^2_{3.8}\oplus \langle \partial_t \rangle $ &
$(1+u^{2}_x)^{-1}$&$\sqrt{1+u^2_x}$ \\ & & \\ $A^2_{3.9}\oplus
\langle \partial_t \rangle \ (q>0)$ & $\frac{\textstyle{\exp(-2q
\arctan u_x)}}{\textstyle{1+u^2_x}}$&$\sqrt{1+u^2_x}\exp(-q\arctan
u_x)$ \\ & & \\ \hline
\end{tabular}
\end{center}

As we have already mentioned, the realizations of the algebras\
$A_{4.i} \ (i=7,8,9)$\ are constructed by extension of the
realizations of the algebra\ $A_{3.3}$\ by an operator\ $e_4$\
of the type (\ref{n4}). Also, while considering the realizations of\
$A_{3.3} = \langle e_1, e_2, e_3 \rangle$, we have taken into
account the isomorphism of this algebra given by\ $e_1 \to e_1, \ e_2 \to
-e_3, \ e_3 \to e_2.$

In this way, we get three inequivalent realizations of the
algebras\ $A_{4.7}$\ and\ $A_{4.9}$
\begin{eqnarray*}
&& A^1_{4.7} = A^1_{3.3}+\hskip -3.8mm \supset \langle t
\partial_t +(x-t) \partial_x +(2u-\frac{1}{2}t^2) \partial_u
\rangle, \\ && A^2_{4.7} = A^2_{3.3}+\hskip -3.8mm \supset \langle
- \partial_t +x \partial_x +2u \partial_u \rangle, \\ && A^1_{4.9}
= A^2_{3.3}+\hskip -3.8mm \supset \langle -(1+t^2) \partial_t
+(q-t)x\partial_x +(2qu-\frac{1}{2}x^2) \partial_u \rangle\ \
(q>0),
\end{eqnarray*}
which are maximal invariance algebras of nonlinear PDEs
(\ref{2.1}). The corresponding forms of the functions\ $F$, $G$\
are given in Table 6.

Next, we have constructed four inequivalent realizations of the
algebra\ $A_{4.8}$\ that are admitted by nonlinear PDEs from the
class (\ref{2.1}):
\begin{eqnarray*}
&& A^1_{4.8} = A^1_{3.3}+\hskip -3.8mm \supset \langle t
\partial_t +qx \partial_x +(1+q)u \partial_u \rangle\ \ (q \in R),
\\ && A^2_{4.8} = A^1_{3.3}+\hskip -3.8mm \supset \langle t
\partial_t +k \partial_x +u \partial_u \rangle\ \ (k \not =0), \\
&& A^3_{4.8} = A^1_{3.3}+\hskip -3.8mm \supset \langle x
\partial_x +u\partial_u +k^{-1}(\partial_t +x\partial_u) \rangle\
\ (k \not =0), \\ && A^4_{4.8} = A^2_{3.3}+\hskip -3.8mm \supset
\langle (1-q)t \partial_t +x \partial_x +(1+q)u \partial_u
\rangle\ \ (|q| \not =1). \\
\end{eqnarray*}

The realizations\ $A^2_{4.8}, A^4_{4.8}$\ are the maximal
invariance algebras of nonlinear heat conductivity equations
belonging to the class of PDEs (\ref{2.1}), and the corresponding
forms of the functions\ $F$, $G$\ are given in Table 6.

The PDE invariant with respect to the realization\ $A^1_{4.8}$\
reduces to the form \be \label{a52} u_t = \lambda |u_x|^{2q-1}
u_{xx} +x +\epsilon |u_x|^q, \ee where\ $\epsilon =0,$ \ $ \lambda
=\pm 1,$\ provided\ $q=0,1$\ and\ $\epsilon=0, \lambda=\pm 1$\ or\
$\epsilon =1, \lambda \not =0$\, if\  $q\not =0,1$.

Investigating the maximal symmetry admitted by (\ref{a52})we find that
for\ $q \not = -\frac{1}{2}$\ the
realization\ $A^1_{4.8}$\ is its maximal invariance algebra. If\ $q
=-\frac{1}{2}$,\ the change of variables
(\ref{h11}) reduces PDE (\ref{a52}) to the Burgers equation
$$v_{\overline{t}}= \lambda v_{\overline{x}\overline{x}}- v
v_{\overline{x}}.$$ The maximal invariance algebra of the Burgers
equation is the semi-direct sum of the algebra\ $sl(2,R)$\ and a
two-dimensional solvable radical.

The equation invariant under the realization\ $A^3_{4.8}$
is
$$u_t = \pm \exp(2k u_x) u_{xx} +x +\epsilon \exp(ku_x),
\ \ k \not =0, \epsilon =0,1.$$
If\ $\epsilon =1$, then the realization\ $A^3_{4.8}$\ is the
maximal invariance algebra of this equation. For\ $\epsilon
=0$,\ the change of variables
$$ \overline{t} = \frac{1}{2k} e^{2kt}, \ \ \ \overline{x} = -x, \
\ v = -2ku +2ktx, \ \ k \not =0,$$
reduces the equation in question to the PDE
$$v_{ \overline{t} } = \pm \exp ( v_{ \overline{x} } )
v_{\overline{x} \overline{x} },$$
which belongs to the class of equations (\ref{a45}).

Finally, after extending the realizations of the Lie algebra\
$A_{3.5}= \langle e_1, e_2, e_3 \rangle $ by an operator\ $e_4$\
of the form (\ref{n4}), we obtain a realization of the algebra\
$A_{4.10}$ of the form
$$A^1_{4.10}= A^3_{3.5} +\hskip -3.8mm \supset \langle 2kt
\partial_t +u \partial_x -x \partial_u \rangle, \ \ k \ge 0,$$
which is the maximal invariance algebra of the equation
$$u_t = \frac{\exp(2k \arctan u_x)}{1+u^2_x} u_{xx} + \beta
|t|^{-\frac{1}{2}} \sqrt{1+u^2_x} \exp(k\arctan u_x), \ \ k \ge 0,
\ \beta \not =0.$$

We give in Table 6 a complete list of inequivalent PDEs of the
form (\ref{2.1}), whose maximal invariance algebras are
non-decomposable four-dimensional solvable Lie algebras.

\newpage
\begin{center}
{\bf Table 6.}\ {\it Invariance of (\ref{2.1}) under
non-decomposable four-dimensional solvable Lie algebras}
\vspace{3mm}

\begin{tabular}{|c|c|c|} \hline
Algebra& $F$&$G$ \\ \hline & & \\ $A^1_{4.2}$ & $\exp(2-q)u_x$&$
\exp(1-q)u_x, \ \ q  \not =0,1$ \\ & & \\ $A^2_{4.2}$ &
$|u_x|^{\frac{1}{q-1}}$&$ (1-q)^{-1} u_x \ln|u_x|, \ \ q  \not
=0,1$ \\ & & \\ $A^1_{4.3}$ & $\exp(-u_x)$&$ \exp(-u_x)$ \\ & & \\
$A^2_{4.3}$ & $1$&$ -u_x \ln|u_x|$ \\ & & \\ $A^1_{4.4}$ & $\exp
u_x$&$ -\frac{1}{2} u^2_x $ \\ & & \\ $A^1_{4.5}$ &
$|u_x|^{\frac{2p-1}{q-p}}$&$ |u_x|^{\frac{q-1}{q-p}}, p<q, p \cdot
q \not =0, \ \ p,q\not =1$ \\ & & \\ $A^2_{4.5}$ & ${\tilde
F}(u_x)$&$0$\\ & & \\ $A^1_{4.6}$ & $\frac{\textstyle{\exp[(q-2p)
\arctan
u_x]}}{\textstyle{1+u^2_x}}$&$\sqrt{1+u^2_x}\exp[(q-p)\arctan
u_x], q\not =p, \ p\ge 0$ \\ & & \\ $A^1_{4.7}$ & $\lambda u_x, \
\lambda \not =0$&$ x +u_x \ln|u_x|$ \\ & & \\ $A^2_{4.7}$ & $\pm
\exp(-2t)$&$ -\frac{1}{2} u^2_x$ \\ & & \\ $A^1_{4.8} \ (q \not
=-\frac{1}{2})$ & $\pm |u_x|^{2q-1}$&$ x$ \\ & & \\ $A^1_{4.8} \
(q \not = 0,1)$ & $\lambda |u_x|^{2q-1}, \ \lambda \not =0$&$
x+|u_x|^q$ \\ & & \\ $A^2_{4.8}$ & $\lambda |u_x|^{-1}, \ \lambda
\not =0$&$ x-k\ln|u_x|, \ k \not =0$ \\ & & \\ $A^3_{4.8}$ & $\pm
\exp(2k u_x)$&$ x+\exp(k u_x), \ k \not =0$ \\ & & \\ $A^4_{4.8} \
(|q| \not =1)$ & $|t|^{\frac{1+q}{1-q}}$&$- \frac{1}{2}u^2_x$ \\ &
& \\ $A^1_{4.9} \ (q>0)$ & $\pm \exp(-2q\arctan t)$&$ \mp
\frac{\textstyle{t \exp(-2q \arctan t)}}{\textstyle{1+t^2}}
-\frac{1}{2} u^2_x$ \\ & & \\ $A^1_{4.10}$ &
$\frac{\textstyle{\exp(2k\arctan
u_x]}}{\textstyle{1+u^2_x}}$&$\beta|t|^{-\frac{1}{2}}\sqrt{1+u^2_x}
\exp(k\arctan u_x),  k\ge 0, \ \beta \not =0$ \\ & & \\ \hline
\end{tabular}
\end{center}

In order to complete the group classification, we have to analyze
nonlinear equations of the form (\ref{2.1}) which admit
five-dimensional invariance algebras. In Section III.3 we have
constructed two nonlinear heat conductivity equations whose
invariance algebras are five-dimensional semi-direct products of
semi-simple and solvable Lie algebras.
According to the results of \cite{ov1} there are three more PDEs
belonging to the class (\ref{a45}), admitting the five-dimensional
algebras
\begin{eqnarray*}
A^2_5 &=& A^2_{4.5}+\hskip -3.8mm \supset \langle t \partial_t -x
\partial_u \rangle; \\ A^3_5 &=& A^2_{4.5}+\hskip -3.8mm \supset
\langle nt \partial_t -u \partial_u \rangle, \ (n \ge -1, n \not
=0); \\ A^4_5 &=& A^2_{4.5}+\hskip -3.8mm \supset \langle nt
\partial_t +u \partial_x -x \partial_u \rangle, \ ( n \ge 0),
\end{eqnarray*}
and there is one PDE admitting the realization\ $A^1_{4.5}$. It is
not difficult to verify that all the algebras\ $A^i_5 \ (i=1,
\ldots, 4)$\ are solvable five-dimensional Lie algebras. Moreover,
the algebra\ $A^1_5$\ is decomposable\ $A^1_5 \sim A_{3.7} \oplus
A_{2.2}$\ and the algebras\ $A^i_5 \ (i=2,3,4)$\ are
non-isomorphic non-decomposable five-dimensional solvable Lie
algebras (see, e.g., \cite{mub1:})
\begin{eqnarray*}
&&A^2_5 \sim A_{5.34} (p=2), A^3_5 \sim A_{5.33} (p=2+n, q=-n)\\
&& A^4_5 \sim A_{5.35} (p=2, q=n).
\end{eqnarray*}

Consequently, the PDEs invariant with respect to the above algebras
are inequivalent. We give in Table 7 a complete list of
nonlinear PDEs of the form (\ref{2.1}) whose maximal invariance
algebras are five-dimensional. \vspace{3mm}

\section{Some conclusions}
\setcounter{equation}{0} \setcounter{thm}{0} \setcounter{rmk}{0}
\setcounter{lemma}{0}

Surprisingly, the number of inequivalent nonlinear PDEs of the general form
under consideration,
and which admit non-trivial symmetry groups is reasonably small.
Summarizing the results of our group classification
of nonlinear heat conductivity equations of the form (\ref{2.1})
we conclude that

\begin{enumerate}

\item{There are two inequivalent nonlinear PDEs (\ref{2.1}), that
admit a one-dimensional invariance algebra.}

\item{There are five inequivalent PDEs (\ref{2.1}) given in Table
1, which are invariant with respect to two-dimensional Lie
algebras. Note that all two-dimensional Lie algebras are
solvable.}

\item{Nonlinear heat conductivity equations (\ref{2.1}) invariant
under three-dimensional Lie algebras (note that a
three-dimensional Lie algebra is either semi-simple or solvable).}
\begin{enumerate}
\item{There are six PDEs (\ref{2.1}) admitting three-dimensional
semi-simple invariance algebras (see Theorems 3.1 and 3.2).}
\item{There are twenty eight equations (\ref{2.1}), given in Tables
2--4, which are invariant with respect to three-dimensional solvable
Lie algebras. }
\end{enumerate}

\item{Nonlinear heat conductivity equations (\ref{2.1}) invariant
under four-dimensional Lie algebras (note that there are no
semi-simple four-dimensional Lie algebras).}
\begin{enumerate}
\item{There are five PDEs (\ref{2.1}) admitting four-dimensional
invariance algebras, that are semi-direct sums of semi-simple and
solvable Lie algebras (see PDEs given at the end of Section
III.3).}
\item{There are thirty equations (\ref{2.1}) given in Tables 5,6,
which are invariant with respect to four-dimensional solvable Lie
algebras. }
\end{enumerate}

\item{Nonlinear heat conductivity equations (\ref{2.1}) invariant
under five-dimensional Lie algebras (note that there are no
semi-simple five-dimensional Lie algebras).}
\begin{enumerate}
\item{There are two PDEs (\ref{2.1}) admitting five-dimensional
invariance algebras, that are semi-direct sums of semi-simple and
solvable Lie algebras (see PDEs given at the end of Section
III.3).}
\item{There are four equations (\ref{2.1}) given in Table 7,
which are invariant with respect to five-dimensional solvable Lie
algebras.}
\end{enumerate}

\end{enumerate}
\vspace{2mm}

{\bf Table 7.}\ {\it Invariance of (\ref{2.1}) under
four-dimensional solvable Lie algebras} \vspace{3mm}

\begin{center}
\begin{tabular}{|c|c|c|} \hline
Algebra& $F$&$G$ \\ \hline & & \\ $A^1_5$&$u^{-2}_x$&\hskip 8mm
$u^{-1}_x \hskip 8mm $\\ & & \\ $A^2_5$&$\exp u_x$&\hskip 8mm$0
\hskip 8mm $\\ & & \\ $A^3_5$&$ u^n_x, \ n \ge -1, \ n \not
=0$&\hskip 8mm $0 \hskip 8mm $ \\ & & \\
$A^4_5$&$\frac{\textstyle{\exp (n \arctan
u_x)}}{\textstyle{1+u^2_x}}, \ n \ge 0$&\hskip 8mm $0\hskip 8mm $
\\ & & \\ \hline
\end{tabular}
\end{center}

We have shown that there are no nonlinear PDEs of the form
(\ref{2.1}) admitting invariance algebras of the dimension higher
than five. Consequently, the classification of
invariant nonlinear heat conductivity equations (\ref{2.1}) presented above  is
complete in the sense that {\bf any} PDEs of the form (\ref{2.1}),
which possess non-trivial Lie symmetry, can be reduced to one of
the canonical forms given above.

Furthermore, we have shown that the results on group
classification of particular equations from the class (\ref{2.1})
obtained in \cite{ah:,zhd99}, \cite{ov1}--\cite{gan} can be
derived from our considerations. That is to say, for each
invariant PDE (\ref{2.1}) obtained in the  papers enumerated
above we can give an invariant equation from our list which is
equivalent to it. The procedure of looking for a corresponding
change of variables is purely algebraic. We start with identifying
the invariance algebra by determining whether it is semi-simple,
or solvable, or a semi-direct sum of semi-simple and solvable
algebras, and then we find the corresponding realizations of the
Lie algebras of the same dimension as the algebra under study.
Comparing the two realizations, it is not difficult to find the
explicit form of the change of variables connecting these
realizations (and, consequently transforming the corresponding
invariant equations  into one another).

One more point is that our classification is in full accordance
with the results of Sokolov \cite{sok88} and Magadeev
\cite{mag93}. These papers discuss, in particular, estimates for the
dimension of the symmetry algebras of evolution PDEs with
one or more spatial variables.

Another important point is the so called quasi-local or non-local
symmetries of nonlinear heat conductivity equations. One can
construct a number of these kind of symmetries as indicated in
\cite{ah1:} and combine them with non-local transformations like
the Legendre, Laplace and B\"acklund transformations
\cite{fs1:,king91}. As mentioned in the introduction, there is an
intriguing possibility of reducing the problem of group
classification of the general second order evolution equation
(\ref{2.1g}) to that of PDE (\ref{2.1}). Moreover, the "singular
points" of this reduction are the quasi-local symmetries of
(\ref{2.1}) which might correspond to usual Lie symmetries of
(\ref{2.1g}). However, these very important questions go beyond
the scope of the present paper and, in fact, can be a basis of a
separate paper.

Let us stress again that it is our belief that the models most
adequately describing real processes should possess the highest
symmetry. That is why the most probable candidates for the roles
of such models are PDEs admitting four and five-dimensional
invariance algebras. The corresponding list of PDEs contain the
well-known equations (like the Burgers equation) and several
principally new equations which certainly deserve further
investigation.

The questions mentioned above are under now study and will be
reported on in future publications.

\bigskip\noindent{\bf Acknowledgements.} R. Zhdanov thanks the Swedish
Natural Sciences Research Council for financial support (grant
number R-RA 521-2373/1999) and the Mathematics Department,
Link\"oping University, for its hospitality and financial support
during his visit to Sweden.

\bigskip
\section*{Appendix 1: solvable Lie algebras.}

\bigskip
\underline{Three-dimensional solvable Lie algebras $(L = \langle
e_1, e_2, e_3 \rangle)$ over ${\bf R}$}

The set of three-dimensional solvable Lie algebras consists of the
following two decomposable Lie algebras:

\bigskip
\begin{eqnarray*}
A_{3.1}&=&  A_1 \oplus A_1\oplus A_1 = 3 A_1; \\ A_{3.2}&=&
A_{2.2} \oplus A_1, \ \ [e_1, e_2] = e_2,
\end{eqnarray*}
and the following eight classes of non-decomposable Lie algebras:
\begin{eqnarray*}
A_{3.3}&:& [e_2, e_3] = e_1;\\ A_{3.4}&:& [e_1, e_3] = e_1, \ \ \
[e_2, e_3] = e_1 +e_2;\\ A_{3.5}&:& [e_1, e_3] = e_1, \ \ \ [e_2,
e_3] = e_2;\\ A_{3.6}&:& [e_1, e_3] = e_1, \ \ \ [e_2, e_3] =
-e_2;\\ A_{3.7}&:& [e_1, e_3] = e_1, \ \ \ [e_2, e_3] = q e_2 \ \
(0 < |q| <1);\\ A_{3.8}&:& [e_1, e_3] = -e_2, \ \ \ [e_2, e_3] =
e_1;\\ A_{3.9}&:& [e_1, e_3] = qe_1-e_2, \ \ \ [e_2, e_3] = e_1+q
e_2, \ (q>0).
\end{eqnarray*}

We note that the algebra $A_{3.3}$ is nilpotent. Note also that we
have $A_{3.i} \ (i=3,4, \ldots, 9)$\ such that $\langle e_1, e_2
\rangle$  $= A_{2.1} =  2 A_1.$

\bigskip
\underline{Four-dimensional solvable Lie algebras $(L = \langle
e_1, e_2, e_3, e_4 \rangle)$ over ${\bf R}$}

\bigskip
Amongst the four-dimensional Lie algebras there are 10
decomposable algebras: $4 A_1 = A_{3.1} \oplus A_1, \ A_{2.2}
\oplus 2 A_1 = A_{2.2} \oplus A_{2.1}, \ A_{2.2} \oplus A_{2.2} =
2 A_{2.2}, A_{3.i} \oplus A_1 \ (i =3,4, \ldots, 9)$; and 10
non-decomposable solvable Lie algebras:
\begin{eqnarray*}
A_{4.1}&:& [e_2, e_4] = e_1, \ \ [e_3, e_4] = e_2;\\ A_{4.2}&:&
[e_1, e_4] = q e_1, \ \ \ [e_2, e_4] = e_2, \ \ [e_3, e_4] = e_2
+e_3, \ \ q \not =0;\\ A_{4.3}&:& [e_1, e_4] = e_1, \ \ \ [e_3,
e_4] = e_2;\\ A_{4.4}&:& [e_1, e_4] = e_1, \ \ \ [e_2, e_4] =
e_1+e_2, \ \ [e_3, e_4] = e_2+e_3;\\ A_{4.5}&:& [e_1, e_4] = e_1,
\ \ \ [e_2, e_4] = q e_2, \ \ [e_3, e_4] = p e_3, \ \ -1 \le p \le
q \le 1, \ \ p \cdot q \not =0;\\ A_{4.6}&:& [e_1, e_4] = q e_1, \
\ \ [e_2, e_4] = p e_2-e_3, \ \ [e_3, e_4] = e_2 +p e_3, \ \ q
\not =0, \ \ p\ge 0;\\ A_{4.7}&:& [e_2, e_3] = e_1, \ \ \ [e_1,
e_4] = 2e_1, \ \ [e_2, e_4] = e_2, \ \ [e_3, e_4] = e_2 +e_3;\\
A_{4.8}&:& [e_2, e_3] = e_1, \ \ \ [e_1, e_4] = (1+q) e_1, \ \
[e_2, e_4] = e_2, \ \ [e_3, e_4] = qe_3, \ \ |q| \le 1;\\
A_{4.9}&:& [e_2, e_3] = e_1, \ \ \ [e_1, e_4] = 2qe_1, \ \ [e_2,
e_4] =q e_2-e_3, \ \ [e_3, e_4] = e_2 +q e_3, \ \ q \ge 0;\\
A_{4.10}&:& [e_1, e_3] = e_1, \ \ \ [e_2, e_3] = e_2, \ \ [e_1,
e_4] =- e_2, \ \ [e_2, e_4] = e_1.
\end{eqnarray*}

\underline{Five-dimensional solvable Lie algebras $(L = \langle
e_1,e_2,
 \ldots, e_5 \rangle)$ over ${\bf R}$}

The set of non-isomorphic five-dimensional Lie algebras is
exhausted by 27 types of decomposable algebras: $ 5 A_1, A_{2.2}
\oplus 3 A_1, 2 A_{2.2} \oplus A_1, A_{3.i} \oplus 2A_1 \ (i=3,4,
\ldots, 8), A_{3.i} \oplus A_{2.2} (i = 3,4, \ldots 8), A_{4.i}
\oplus A_1 \ (i =1, \ldots, 10)$; and 39 non-decomposable solvable
algebras:
\begin{eqnarray*}
A_{5.1}&:& [e_3, e_5] = e_1, \ \ [e_4, e_5] = e_2;\\ A_{5.2}&:&
[e_2, e_5] =  e_1, \ \ \ [e_3, e_5] = e_2, \ \ [e_4, e_5] =e_3;\\
A_{5.3}&:& [e_2, e_4] = e_3, \ \ \ [e_2, e_5] = e_1, \ \  [e_4,
e_5] =e_2;\\ A_{5.4}&:& [e_2, e_4] = e_1, \ \ \ [e_3, e_5] =
e_1;\\ A_{5.5}&:& [e_3, e_4] = e_1, \ \ \ [e_2, e_5] = e_1, \ \
[e_3, e_5] =e_2;\\ A_{5.6}&:& [e_3, e_4] = e_1, \ \ \ [e_2, e_5] =
e_1, \ \ [e_3, e_5] = e_2, \ \ [e_4, e_5] = e_3;\\ A_{5.7}&:&
[e_1, e_5] = e_1, \ \ \ [e_2, e_5] = p e_2, \ \ [e_3, e_5] = q
e_3, \\ && [e_4, e_5] = r e_4, \ \ -1 \le r \le q \le p \le 1, \
rpq \not =0;\\ A_{5.8}&:& [e_2, e_5] = e_1, \ \ \ [e_3, e_5] =
e_3, \ \ [e_4, e_5] =p e_4, \ \ 0< |p| \le 1;\\ A_{5.9}&:& [e_1,
e_5] = e_1, \ \ \ [e_2, e_5] = e_1+e_5, \ \ [e_3, e_5] =p e_3, \ \
[e_4, e_5] = q e_4, \ \ 0 \not = q \le p;\\ A_{5.10}&:& [e_2, e_5]
= e_1, \ \ \ [e_3, e_5] = e_2, \ \ [e_4, e_5] = e_4; \\
A_{5.11}&:& [e_1, e_5] = e_1, \ \ \ [e_2, e_5] = e_1+e_2, \ \
[e_3, e_5] =e_2+ e_3, \ \ [e_4, e_5] = p e_4, \ \ p  \not = 0;\\
A_{5.12}&:& [e_1, e_5] = e_1, \ \ \ [e_2, e_5] = e_1+e_2, \ \
[e_3, e_5] = e_2+e_3, \ \ [e_4, e_5] = e_3+e_4; \\ A_{5.13}&:&
[e_1, e_5] = e_1, \ \ \ [e_2, e_5] = p e_2, \ \ [e_3, e_5] = q
e_3-r e_4, \\ && [e_4, e_5] = q e_4+r e_3, \ \ |p|\le 1, \ \ p
\cdot r \not =0, \ \ q \ge 0;\\ A_{5.14}&:& [e_2, e_5] = e_1, \ \
\ [e_3, e_5] = p e_3-e_4, \ \ [e_4, e_5] =  e_3+p e_4, \ \ p \ge
0;\\ A_{5.15}&:& [e_1, e_5] = e_1, \ \ \ [e_2, e_5] =  e_1+e_2, \
\ [e_3, e_5] =p e_3, \\ && [e_4, e_5] =  e_3+p e_4, \ -1 \le p \le
1;\\ A_{5.16}&:& [e_1, e_5] = e_1, \ \ \ [e_2, e_5] = e_1+ e_2, \
\ [e_3, e_5] = p e_3-q e_4, \\ && [e_4, e_5] = q e_3+p e_4, \ \
p\ge 0, \ \ q \not =0;\\ A_{5.17}&:& [e_1, e_5] = pe_1-e_2, \ \ \
[e_2, e_5] = e_1+ p e_2, \ \ [e_3, e_5]= q e_3-r e_4, \\ && [e_4,
e_5] = r e_3+q e_4, \ \  r \not =0, \ \ p, q \in {\bf R};\\
A_{5.18}&:& [e_1, e_5] = p e_1- e_2, \ \ \ [e_2, e_5] = e_1+ p
e_2, \ \ [e_3, e_5]= e_1+p e_3- e_4, \\ && [e_4, e_5] = e_2+e_3-p
e_4, \ \ p \in {\bf R};\\ A_{5.19}&:& [e_2, e_3] =  e_1, \ \ \
[e_1, e_5] = (1+p)e_1, \ \ [e_2, e_5]= e_2, \ \ [e_3, e_5] = p
e_3, \\ && [e_4, e_5] = q e_4, \ p \in {\bf R}, \ q \not =0;\\
A_{5.20}&:& [e_2, e_3] =  e_1, \ \ \ [e_1, e_5] = (1+p)e_2, \ \
[e_2, e_5]= e_2, \ \ [e_3, e_5] = p e_3, \\ && [e_4, e_5] = e_1
+(1+p)e_4, \ \ p, q \in {\bf R};\\ A_{5.21}&:& [e_2, e_3] = e_1, \
\ \ [e_1, e_5] = 2 e_1, \ \ [e_2, e_5]= e_2+ e_3, \ \ [e_3, e_5] =
e_3+ e_4, \ \ [e_4, e_5] = e_4;\\ A_{5.22}&:& [e_2, e_3] = e_1, \
\ \ [e_2, e_5] = e_3, \ \ [e_4, e_5] =  e_4;\\ A_{5.23}&:& [e_2,
e_3] =  e_1, \ \ \ [e_1, e_5] = 2 e_1, \ \ [e_2, e_5]= e_2+e_3, \\
&& [e_3, e_5] =  e_3, \ \ [e_4, e_5] = p e_4,
 \ \ p\not =0;\\
A_{5.24}&:& [e_2, e_3] = e_1, \ \ \ [e_1, e_5] = 2 e_1, \ \ [e_2,
e_5]= e_2+ e_3, \\ && [e_3, e_5] = e_3, \ \ [e_4, e_5] = \epsilon
e_1+2 e_4, \ \ \epsilon = \pm 1;\\ A_{5.25}&:& [e_2, e_3] = e_1, \
\ \ [e_1, e_5] = 2 p e_1, \ \ [e_2, e_5] = p e_2+ e_3, \ \ [e_3,
e_5] = -e_2 +p e_3, \\ && [e_4, e_5] = q e_4, \ \ p \in {\bf R}, \
\ q \not = 0;\\ A_{5.26}&:& [e_2, e_3] = e_1, \ \ \ [e_1, e_5] = 2
p e_1, \ \ [e_2, e_5] = p e_2+ e_3, \ \ [e_3, e_5] = -e_2 +p e_3,
\\ && [e_4, e_5] = \epsilon e_1+ 2 p e_4, \ \ \epsilon = \pm 1 , \
\ p \in {\bf R};\\ A_{5.27}&:& [e_2, e_3] = e_1, \ \ \ [e_1, e_5]
= e_1, \ \ [e_3, e_5] = e_3+ e_4, \ \ [e_4, e_5] = e_1+e_4;\\
A_{5.28}&:& [e_2, e_3] =  e_1, \ \ \ [e_1, e_5] = (1+p)e_1, \ \
[e_2, e_5]= p e_2, \\ && [e_3, e_5] =  e_3+e_4, \ \ [e_4, e_5] =
e_4 , \ \ p\in {\bf R};\\ A_{5.29}&:& [e_2, e_3] =  e_1, \ \ \
[e_1, e_5] = e_1, \ \ [e_2, e_5]=  e_2, \ \ [e_3, e_5] =  e_4;\\
A_{5.30}&:& [e_2, e_4] =  e_1, \ \ \ [e_3, e_4] =  e_2, \ \ [e_1,
e_5]= (2+p)e_1, \ \ [e_2, e_5] =  (1+p)e_2, \\ && [e_3, e_5] = p
e_3,
 \ \ [e_4, e_5] = e_4 , p\in {\bf R};\\
A_{5.31}&:& [e_2, e_4] =  e_1, \ \ \ [e_3, e_4] = e_2, \ \ [e_1,
e_5]= 3 e_1, \\ && [e_2, e_5] = 2 e_2, \ \ [e_3,e_5] = e_3, \ \
[e_4, e_5] = e_3+e_4;\\ A_{5.32}&:& [e_2, e_4] =  e_1, \ \ \ [e_3,
e_4] = e_2, \ \ [e_1, e_5]=  e_1, \ \ [e_2, e_5] =  e_2, \\ &&
[e_3,e_5] = p e_1 + e_3, \ \ p \in {\bf R};\\ A_{5.33}&:& [e_1,
e_4] =  e_1, \ \ \ [e_3, e_4] = p e_3, \ \ [e_2, e_5]=  e_2, \\ &&
[e_3, e_5] =   q e_3, \ \ p, q \in {\bf R},  \ \ p^2+q^2 \not
=0;\\ A_{5.34}&:& [e_1, e_4] =  p e_1, \ \ \ [e_2, e_4] = e_2, \ \
[e_3, e_4]=  e_3, \ \ [e_1, e_5] =  e_1, \ \ [e_3,e_5] =  e_2, \ \
p \in {\bf R};\\ A_{5.35}&:& [e_1, e_4] =  p e_1, \ \ \ [e_2, e_4]
=  e_2, \ \ [e_3, e_4]= e_3, \ \ [e_1, e_5] = q e_1, \\ && [e_2,
e_5] =-e_3, \ \ [e_3, e_5] = e_2, p, q \in {\bf R},  \ \ p^2+q^2
\not =0;\\ A_{5.36}&:& [e_2, e_3] =  e_1, \ \ \ [e_1, e_4] = e_1,
\ \ [e_2, e_4]= e_2, \ \ [e_2, e_5] = - e_2, \ \ [e_3,e_5] =
e_3;\\ A_{5.37}&:& [e_2, e_3] =  e_1, \ \ \ [e_1, e_4] = 2 e_1, \
\ [e_2, e_4]= e_2, \\ && [e_3, e_4] = e_3, \ \ [e_2,e_5] = - e_3,\
\ [e_3,e_5] = e_2;\\ A_{5.38}&:& [e_1, e_4] =  e_1, \ \ \ [e_2,
e_5] = e_2, \ \ [e_4, e_5]= e_3;\\ A_{5.39}&:& [e_1, e_4] =  e_1,
\ \ \ [e_2, e_4] =  e_2, \ \ [e_1, e_5]= -e_2, \ \ [e_2, e_5] =
e_1, \ \ [e_4,e_5] =  e_3.
\end{eqnarray*}

\bigskip
\section*{Appendix 2: Lie algebras which are semi-direct sums of
semi-simple and solvable algebras.}

\begin{enumerate}
\item \underline{Lie algebras of dimensions 5 and 6.}
\begin{eqnarray*}
sl(2,{\bf R}) \subset \hskip -3.8mm + A_{2.1}&:& [e_1, e_4] = e_4,
\ \ [e_2, e_5] = e_4, \ [e_3, e_4] = e_5, \ [e_1, e_5] = -e_5;\\
&&  \\ so(3) \subset \hskip -3.8mm + A_{3.1}&:& [e_1, e_5] = e_6,
\ \ [e_2, e_4] = -e_6, \ [e_3, e_4] = e_5, \ [e_1, e_6] = -e_5,\\
&& [e_2, e_6] = e_4, \ \ [e_3, e_5] = -e_4;\\ &&  \\ sl(2,{\bf R})
\subset \hskip -3.8mm + A_{3.i}&:& [e_1, e_4] = e_4, \ \ [e_2,
e_5] = e_4, \\ i=3, A_{3.3} = \langle e_6, e_4, e_5 \rangle &&
[e_3, e_4] = e_5, \ [e_1, e_5] = -e_5,\\ i=5, \  A_{3.5} = \langle
e_4, e_5, e_6 \rangle;&& \\ &&  \\ sl(2,{\bf R}) \subset \hskip
-3.8mm + A_{3.1}&:& [e_1, e_4] = 2 e_4, \ [e_2, e_5] = 2 e_4, \
[e_3, e_4] = e_5, \\ && [e_1, e_6] = -2 e_6, \ [e_2, e_6] = e_5, \
[e_3, e_5] = 2 e_6.
\end{eqnarray*}
\item \underline{Lie algebras of dimension 7.}
\begin{eqnarray*}
so(3) \subset \hskip -3.8mm + A_{4.5} (p=q=1)&:& [e_1, e_5]=e_6, \
[e_2, e_4] = -e_6, \ [e_3, e_4] = e_5, \\ && [e_1, e_6] = -e_5, \
[e_2, e_6] = e_4, \ [e_3, e_5] = -e_4;\\ &&  \\ so(3) \subset
\hskip -3.8mm + 4 A_{1} \hskip 15mm &:& [e_1, e_4]=\frac{1}{2}
e_7, \ [e_2, e_4] = \frac{1}{2}e_5, \ [e_3, e_4] = \frac{1}{2}
e_6, \\ && [e_1, e_5] = \frac{1}{2}e_6, \ [e_2, e_5] =
-\frac{1}{2} e_4, \ [e_3, e_5] = -\frac{1}{2} e_7,\\ && [e_1, e_6]
= -\frac{1}{2} e_5, \ [e_2, e_6] = \frac{1}{2} e_7, \ [e_3, e_6] =
-\frac{1}{2} e_4, \\ && [e_1, e_7] = -\frac{1}{2} e_4, \ [e_2,
e_7] = -\frac{1}{2} e_6, \ [e_3, e_7] = \frac{1}{2} e_5;\\ && \\
sl(2,{\bf R}) \subset \hskip -3.8mm + A_{4.i} &:& [e_1, e_4] =
e_4, \ [e_2, e_5] = e_4, \\
 i=5: A_{4.5} \ (q=1), && [e_3, e_4] = e_5, \ [e_1, e_5] = -e_5,\\
i=8: A_{4.8} (q=1), && \\
 A_{4.8} = \langle e_6, e_4, e_5, e_7
\rangle;&& \\ && \\ sl(2,{\bf R}) \subset \hskip -3.8mm +
A_{4.5}&:& [e_1, e_4] = 2 e_4, \ [e_2, e_5] = 2 e_4, \ [e_3, e_4]
= e_5, \\ A_{4.5} (p=q=1) &&[e_1, e_6] = -2 e_6, \ [e_2, e_6] =
e_5, \ [e_3, e_5] = 2 e_6;\\ && \\ sl(2,{\bf R}) \subset \hskip
-3.8mm + 4 A_1 &:& [e_1, e_4] = 3 e_4, \ [e_2, e_5] = 3 e_4, \
[e_3, e_4] = e_5, \\ && [e_1, e_5] = e_5, \ [e_2, e_6] = 2 e_5, \
[e_3, e_5] = 2 e_6, \\ && [e_1, e_6] = -e_6, \ [e_2, e_7] = e_6, \
[e_3, e_6] = 3 e_7, \ [e_1, e_7] = -3 e_7;\\ && \\ sl(2,{\bf R})
\subset \hskip -3.8mm + 4 A_1 &:& [e_1, e_4] = e_4, \ [e_2, e_5] =
e_4, \ [e_3, e_4] = e_5,  \ [e_1, e_5] = -e_5,\\ && [e_1, e_6] =
e_6, \ [e_2, e_7] = e_6, \ [e_3, e_6] = e_7, \ [e_1, e_7] = -e_7.
\end{eqnarray*}
\item \underline{Lie algebras of dimension 8.}
\begin{eqnarray*}
so(3) \subset \hskip -3.8mm + A_{5.7}&:& [e_1, e_5] = e_6, \ [e_2,
e_4] = -e_6, \ [e_3, e_4] = e_5,\\ A_{5.7} (p=q=1) && [e_1, e_6] =
-e_5, \ [e_2, e_6]=e_4, \ [e_3, e_5] = -e_4;\\ && \\ so(3) \subset
\hskip -3.8mm + A_{5.i}&:& [e_1, e_4] = \frac{1}{2} e_7, \ [e_2,
e_4] = \frac{1}{2} e_5, \\
 i=4: A_{5.4} = \langle e_8, e_4, e_7, e_5, e_6 \rangle, && [e_3, e_4]
= \frac{1}{2} e_6, \ [e_1, e_5] = \frac{1}{2} e_6,\\
 i=7: A_{5.7} (p=q=r=1),&& [e_2, e_5] = -\frac{1}{2} e_4, \ [e_3,
e_5] = -\frac{1}{2}e_7,\\
 i=17: A_{5.17} \ (p=q, \ r=1), && [e_1, e_6] = -\frac{1}{2} e_5, \
[e_2, e_6] = \frac{1}{2} e_7,\\
 A_{5.17} = \langle e_4, e_6, e_5, e_7, e_8 \rangle,&& [e_3, e_6] =
-\frac{1}{2} e_4, \ [e_1, e_7] = -\frac{1}{2} e_4,\\ && [e_2, e_7]
= -\frac{1}{2} e_6, \ [e_3, e_7] = \frac{1}{2} e_5;\\  && \\ so(3)
\subset \hskip -3.8mm + 5 A_1&:& [e_1, e_4] = \frac{1}{2} e_7, \
[e_1, e_5]= -\frac{1}{2} e_6, \\ &&  [e_1, e_6] = 2 e_5 -e_8, \
[e_1, e_7] = -2 e_4, \\ && [e_1, e_8] = 3 e_6, \ [e_2, e_4] =
\frac{1}{2} e_6, \\ &&  [e_2, e_5] = \frac{1}{2} e_7, \ [e_2, e_6]
= -2 e_4, \\ && [e_2, e_7] = -2 e_5-e_8, \ [e_2, e_8] = 3 e_7, \\
&&  [e_3, e_4] = 2 e_5, \ [e_3, e_5] = -2 e_4, \\ && [e_3, e_6] =
e_7, \ [e_3, e_7] = - e_6;\\ && \\ sl(2,{\bf R}) \subset \hskip
-3.8mm +  A_{5.i}&:& [e_1, e_4] = e_4, \ [e_2, e_5]= e_4, \\ &&
[e_3, e_4] =  e_5, \ [e_1, e_5] = - e_5; \\
 i=4:A_{5.4} = \langle e_8, e_4, e_6, e_5, e_7 \rangle, && \\
 i=7,8: A_i (p=1),&& \\
A_{5.8} = \langle e_6, e_7, e_4, e_5, e_8 \rangle, && \\
 i=9: \cong A_{5.9}, && \\
 i=13,19,20: A_{5.i} \ (p=1), && \\
 A_{5.i} \ (i=19,20)= \langle e_6,e_4, e_5, e_7, e_8 \rangle, && \\ && \\
sl(2,{\bf R}) \subset \hskip -3.8mm +  A_{5.7}&:& [e_1, e_4] =
2e_4, \ [e_2, e_5]= 2e_4, \ [e_3, e_4] =  e_5, \\ && [e_1, e_5] =
-2 e_6; \ \  [e_2, e_6 ] = e_5, \ \ [e_3, e_5] = 2 e_6;\\ && \\
sl(2,{\bf R}) \subset \hskip -3.8mm +  A_{5.i}&:& [e_1, e_4] =
e_4, \ [e_2, e_5]= e_4, \\
 i=4:A^\epsilon_{5.4},&&  [e_3, e_4] =  e_5, \ [e_1, e_5] = - e_5, \\
 i=1: A_{5.1},&&  [e_1, e_6] =  e_6, \ [e_2, e_7] =  e_6, \\
 i=3:\cong  A_{5.3}, &&  [e_3, e_6] = e_7, \ [e_1, e_7] = -e_7, \\
 i=15: A_{5.15}\ (p=1) && \\
 i=7: A_{5.7}\ (p=q=1, \ -1 \le r \le 1) && \\
 i=17: A_{5.17}\ (p=q, \ r=1, p \ge 0) && \\
 A_{5.i} \ (i=7,17) = \langle e_4, e_6, e_5, e_7, e_8 \rangle;&& \\ && \\
sl(2,{\bf R}) \subset \hskip -3.8mm +  A_{5.i}&:& [e_1, e_4] =
3e_4, \ [e_2, e_5]= 3e_4, \\ && [e_3, e_4] =  e_5, \ [e_1, e_5] =
e_5; \\ i=4: \cong A_{5.4}&& [e_2, e_6] = 2 e_5, \ [e_3, e_5] = 2
e_6, \ [e_1, e_6] = - e_6, \\ i=7: A_{5.7} (p=q=r=1) && [e_2,
e_7]= e_6, \ [e_3, e_6] = 3 e_7, \ [e_1, e_7] = -3 e_7; \\ && \\
sl(2,{\bf R}) \subset \hskip -3.8mm +  5 A_{1}&:& [e_1, e_4] =
4e_4,
 \ [e_2, e_5]= 4 e_4, \\ && [e_3, e_4] =  e_5, \ [e_1, e_5] = 2 e_5; \\
&&  [e_2, e_6] = 3 e_5, \ [e_3, e_5]= 2 e_6, \\ && [e_1, e_7] = -2
e_7, \ [e_2, e_7] = 2 e_6, \\ &&  [e_3, e_6] = 3 e_7,\ [e_1, e_8]=
-4 e_8, \\ && [e_2, e_8] =  e_7, \ [e_3, e_7] = 4 e_8;\\ &&
\\ sl(2,{\bf R}) \subset \hskip -3.8mm +  5 A_{1}&:& [e_1, e_4] =
2 e_4, \ [e_2, e_5]= 2 e_4, \\ && [e_3, e_4] =  e_5, \ [e_1, e_6]
= -2 e_6; \\ &&  [e_2, e_6] =  e_5, \ [e_3, e_5]= 2 e_6, \\ &&
[e_1, e_7] =   e_7, \ [e_2, e_8] =  e_7, \\ &&  [e_3, e_7] =  e_8,
\ [e_1, e_8]= - e_8.
\end{eqnarray*}
\end{enumerate}

In giving the type of the radicals, we have followed the rule that
the bases of the radicals consist of the operators
$e_4,\ldots,e_m,$ whenever the basis is not given explicitly;
where it is given explicitly, then the basis operators are ordered
as in the corresponding solvable algebra. For instance, in the
algebra $sl(2,{\bf R}) \subset \hskip -3.8mm + A_{5.17}$ we have
written $A_{5.17} = \langle e_4, e_6, e_5, e_7, e_8 \rangle.$ This
means that the basis operators satisfy the commutation relations
which define the algebra $A_{5.17}$ given in the list of solvable
algebras. To obtain the commutation relations for the algebra
$A_{5.17},$ we replace the operators $e_4, e_5, e_6, e_7, e_8$ as
follows:
\[
e_4\to e_1, \ \ e_6 \to e_2, \ \ e_5 \to e_3, \ \ e_7 \to e_4, \ \
e_8 \to e_5.
\]
Furthermore, for the five-dimensional radicals $N=\langle
e_4,e_5,e_6,e_7,e_8\rangle$ we use the notation
\begin{eqnarray*}
\cong A_{5.9}&:& [e_4, e_8] = e_4, \ \ [e_5, e_8] = e_5, \ \ [e_6,
e_8] = p e_6, \\ && [e_7, e_8] = e_6+p e_7, \ p\not =0;\\
A^\epsilon_{5.4}&:& [e_4, e_8] = e_8, \ \ [e_6, e_7] = \epsilon
e_8, \ \ \epsilon = \pm 1;\\ \cong A_{5.3}&:& [e_6, e_8] = e_4, \
\ [e_7, e_8] = e_5, \ \ [e_6, e_7] =  e_8; \\ \cong A_{5.4}&:&
[e_4, e_7] = e_8, \ \ [e_5, e_6] = -3 e_8.
\end{eqnarray*}

\end{document}